\documentclass{aa} 
\usepackage{graphicx, amsmath, color, mathrsfs}
\graphicspath{{figures_finales/}}

\usepackage{natbib,twoopt}
\bibpunct{(}{)}{;}{a}{}{,} 

\usepackage{txfonts}
\newcommand{\mute}[1]{}
\begin{document}
   \title{Variance of the Galactic nuclei cosmic ray flux}

   \author{G. Bernard \inst{1} \and T. Delahaye\inst{2}
     \and
     P. Salati \inst{1} \and R. Taillet \inst{1}}
   
   \institute{LAPTh, Univ. de Savoie, CNRS. B.P. 110. Annecy-le-Vieux F-74941. France
   \thanks{Laboratoire d'Annecy-le-Vieux de Physique Th\'eorique, UMR5108}
 \and
Instituto de F\'isica Te\'orica UAM/CSIC, Universidad Aut\'onoma de
Madrid, Cantoblanco, 28049 Madrid, Spain}

\date{Received ; accepted ;\\Preprint numbers LAPTh-019/12 and IFT-UAM/CSIC-12-34}

 
  \abstract
   {Measurements of cosmic ray fluxes by the PAMELA and CREAM 
     experiments show unexpected spectral features between 200 GeV and
     100 TeV. They could be due to the presence of nearby and young
     cosmic ray sources. This can be studied in the myriad model,
     in which cosmic rays diffuse from point-like instantaneous
     sources located randomly throughout the Galaxy.}
   {To test this hypothesis, one must compute the flux due to a
     catalog of local
     sources, but also the error bars associated to this
     quantity. This turns out not to be as straightforward as it
     seems, as the standard deviation is infinite when computed for
     the most general statistical ensemble. The goals of this paper are to provide a method to associate
     error bars to the flux measurements which has a clear statistical
     meaning, and to explore the relation between the myriad model and
   the more usual source model based on a continuous distribution.}
   {To this end, we show that the quantiles of the flux distribution
     are well-defined, even though the standard deviation is
     infinite. They can be used to compute 68 \% confidence levels,
     for instance. We also use the fact that local sources have known
     positions and ages to reduce the statistical ensemble from which
     random sources are drawn in the myriad model.}
   {We present a method to evaluate meaningful error bars for the flux
     obtained in the myriad model. In this context, we also discuss the status of the
     spectral features observed in the proton flux by CREAM and PAMELA.}
   {}


   \maketitle


\section{Introduction}

In the 1 GeV to 100 TeV energy range, cosmic ray (CR) nuclei that reach
the Earth have a Galactic origin. They were accelerated in sources,
the nature of which is still a subject of discussion and research,
supernova driven shock waves being a good candidate. They subsequently
reach the Earth, after diffusing in the Galactic magnetic field.
The exact locations and ages of the supernova explosions are not known,
and properties of Galactic cosmic rays are often studied under the assumption
that sources form a distribution which is continuous, in space as in time.
In the standard model of CR propagation, sources are modelled as a jelly
which extends inside the Galactic disk and steadily injects particles in
the interstellar medium (ISM). This picture has proven to be quite
successfull so far.
The CR spectra are actually dominated by sources which are distant from Earth
(for instance see Fig.~8 in~\citep{2004ApJ...609..173T}) and the continuous
hypothesis is expected to be valid.
On the other hand, this approach may fail, when one considers the high-energy
end of the CR spectra, mainly produced by young sources. The flux at high energy
could well be dominated by a handful of individual sources, the positions and ages
of which have a strong influence on the spectrum. This is of tremendous importance
as the latest measurements, such as those of
CREAM~\citep{2011ApJ...728..122Y} or
PAMELA~\citep{2011Sci...332...69A}, exhibit spectral features which are
difficult to understand in a continuous and steady-state modelling of CR
sources.

In this article, we study the relation between the mean-field approach and
the so-called ``myriad model'' (\citep{2003ApJ...592..161H}) where CR primary
species, such as protons or helium nuclei, diffuse from point-like sources
distributed randomly in space and time. In the myriad model, the flux of cosmic
rays is a random variable. When averaged over many realisations of the possible
positions and ages of the sources, the myriad model gives the same results as the
continuous model, as expected.

However, we expect fluctuations from the average flux, from one realization to the
other. If these fluctuations are small, then the myriad model predictions of the flux
are robust and similar to that of the continuous model. In that case, the spectral
features seen by CREAM and PAMELA should be explained within the continuous model.
Invoking a distorted spectral injection spectrum at the sources
(see for instance \citet{2011PhRvD..84d3002Y}) or a non-standard
behaviour of the Galactic diffusion coefficient with energy are two possible, although
unlikely, solutions. On the other hand, if the statistical fluctuations of the myriad
model are large, it could be easier, and to some extent more natural, to explain the
CREAM and PAMELA anomalies in terms of the particular history and morphology of the
population of sources within which we live. If so, there is little reason to trust
anymore the continuous model, on which most of CR studies are nevertheless based.
The ``real'' sources could generate CR fluxes at the Earth quite different from
those predicted in the conventional approach. Unfortunately, we do not know where
and when the ``real'' supernova explosions took place. We are reduced to rely on
statistics in order to gauge how probable are CR fluxes far from the values predicted
by the continuous model.

The amplitudes of the statistical fluctuations of the myriad model have actually been
recently studied through the standard deviation of the random variable associated to
the CR proton flux (see e.g.~\citep{2011arXiv1105.4521B}). This quantity turns out te
be infinite, when computed without precautions. This unexpected result is, a priori,
a severe blow against the mean-field approach on which most of the public codes of CR
Galactic transport, such as \textsc{galprop\footnote{http://galprop.stanford.edu/}},
\textsc{dragon\footnote{http://www.desy.de/~maccione/DRAGON/}} or
\textsc{usine\footnote{http://lpsc.in2p3.fr/usine}}, are based. These codes would
have to be entirely modified.
Abandoning the conventional CR model would have also drastic consequences on the
problem of the astronomical dark matter (DM). The nature of this component, which
contributes substantially to the mass budget of the universe, is still an unresolved
issue. Weakly interacting massive particles (WIMP) have been invoked as a plausible
solution. These putative DM species are expected to annihilate inside the Galactic
DM halo, producing in particular rare antimatter cosmic rays such as positrons and
antiprotons. Distortions in the spectra of these particles would be an indirect probe
of the presence of WIMPs inside the Milky Way. Theoretical predictions have been based
so far on the continuous CR model. They need to be entirely revisited should the myriad
model replace the conventional scheme.

This article is devoted to the myriad model and to the issue arising from the infinite
variance of the primary CR flux. We suggest two ways to solve this problem.
First, we show that the statistical distribution of the flux is such that the confidence
intervals are finite and well-defined, even though the variance is infinite. We study
how the width of these intervals depends on the CR propagation model.
Second, we notice that the divergence of the variance is due to the sources that are very
close to the Earth and very young. In our Galaxy, theses objects are known actually, and
the region in which we have catalogs of the sources should be excluded from the statistical
ensemble of the myriad model, and rather treated as known sources. This paper explores the
consequences of this approach.

\section{Propagator for diffusive propagation from discrete sources}

%
%
After their acceleration by supernova driven shock waves, primary
cosmic ray (CR) nuclei subsequently propagate through the Galactic
magnetic fields and bounce off their irregularities -- the Alfv\'en
waves. The resulting particle transport is well described by space
diffusion with a coefficient
\begin{equation}
D(E) = D_{0} \; \beta \; \mathscr{R}^{\delta} \;\; ,
\label{eq:dif_coef}
\end{equation}
which increases as a power law with the rigidity
${\mathscr R} \equiv {p}/{Z}$ of the particle.
In addition, since these scattering centers move  with a velocity
$V_{a} \sim$ 20 to 100 km s$^{-1}$, a second order Fermi mechanism
is  responsible for some diffusive re-acceleration which turns out
to be mostly relevant at low energy, below a few GeV. Since we are
interested here in the excess in CR protons and helium nuclei measured
by CREAM, and more recently by PAMELA, we will disregard diffusive
re-acceleration as well as energy losses which are also negligible
at high energy.
In addition to space diffusion, Galactic convection wipes cosmic
rays away from the Galactic disk with a velocity
$V_{C} \sim$ 5 to 15 km s$^{-1}$.
The master equation for the space and energy number density
$\psi \equiv {dn_{p}}/{dT_{p}}$ of, say, CR protons with kinetic energy
$T_{p}$ takes into account space diffusion and Galactic convection
\begin{equation}
{\displaystyle \frac{\partial \psi}{\partial t}} \, + \,
\partial_{z} \! \left( V_{C} \, \psi \right) \, - \,
D \, \triangle \psi \, = \, q_{\rm eff} =
q_{\rm acc} \, - \, q_{\rm col} \;\; .
\label{master_equation}
\end{equation}
This equation applies to any nuclear species -- protons and helium
nuclei in particular -- as long as the effective rate of production
$q_{\rm eff}$ is properly accounted for.
This general framework, summarized in Eq.~(\ref{master_equation}), is
generally implemented within an axisymmetric two-zone model which is
extensively discussed in~\citet{2001ApJ...555..585M} or \citet{2001ApJ...563..172D},
and whose salient features are briefly recalled here.

%
%
%
The CR diffusive halo (DH) is pictured as a thick wheel which matches
the circular structure of the Milky Way. The Galactic disk of stars and
gas -- where primary CR protons and helium nuclei are accelerated -- lies
in the middle. It extends radially 20 kpc from the center and has a
half-thickness $h$ of 100 pc. Confinement layers, where cosmic rays are
trapped by diffusion, lie above and beneath that disk. The interGalactic
medium starts at the vertical boundaries $z = \pm L$ as well as beyond a
radius of $r = R_{\rm gal} =$ 20~kpc. Notice that the half-thickness $L$
of the DH is not known and possible values range from 1 to 15 kpc.
The diffusion coefficient $D$ is taken to be the same everywhere while
the convective velocity is exclusively vertical, with component
\begin{equation}
V_{C}(z) = V_{C} \, {\rm sign}(z) \;\; .
\end{equation}
The Galactic wind -- produced by the bulk of the disk stars like the Sun
-- drifts away from them along the vertical directions, hence the particular
form assumed here for $V_{C}$.
The effective production term $q_{\rm eff}$ takes into account the
release of primary CR nuclei described by the positive term $q_{\rm acc}$
as well as a negative contribution that accounts for their interactions
with the interstellar gas of the disk with rate $q_{\rm col}$.
In most propagation codes, the injection
of cosmic rays is described by a smooth function of space and is constant
in time. In this article, we treat supernova explosions as point-like events.
The production rate of CR nuclei through acceleration is given by
\begin{equation}
q_{\rm acc}(\mathbf{x}_{S},t_{S}) =
{\displaystyle \sum_{n \in \mathscr{P}}} \,\, q_n \,\,
\delta^{3}(\mathbf{x}_{S} - \mathbf{x}_{n}) \,
\delta(t_{S} - t_{n}) \;\; ,
\label{q_acc_discret}
\end{equation}
where each source $i$ belonging to the population $\mathscr{P}$ of supernovae
contributes a factor $q_{i}$ at position $\mathbf{x}_{i}$ and time $t_{i}$.
The collisions of CR nuclei on the hydrogen and helium atoms of the
interstellar medium (ISM) tend to deplete the high energy regions of
their spectra. In the case of CR protons, which we will use throughout
this section as an illustration, they act as a negative source term for
the energy bin located at $T_{p}$, with amplitude
\begin{equation}
q_{\rm col}(\mathbf{x}_{S},t_{S}) = 2 \, h \, \delta(z_{S}) \;
\Gamma_{p} \; \psi(\mathbf{x}_{S},t_{S}) \;\; .
\end{equation}
The ISM is distributed within an infinitely thin disk, hence the
presence of a $\delta(z_{S})$ function in the previous expression. The
collision rate may be expressed as
\begin{equation}
\Gamma_{p} = v_{p} \;
\left(
\sigma_{p{\rm H}} \, n_{\rm H} \; + \; \sigma_{p{\rm He}} \, n_{\rm He}
\right) \;\; ,
\end{equation}
where the densities $n_{\rm H}$ and $n_{\rm He}$ have been respectively
averaged to $0.9$ and $0.1$ cm$^{-3}$. The total proton-proton cross section
$\sigma_{p{\rm H}}$ has been parameterized according to \cite{Nakamura}
while $\sigma_{p{\rm He}}$ is related to $\sigma_{p{\rm H}}$ by the
\citet{2007NIMPB.254..187N} scaling factor $4^{2.2/3}$. Similar scaling factors
have been used in order to derive the CR helium nuclei collision cross
sections from the proton case.

%
%
%

\subsection{Solution of the diffusion equation}

Our description of the propagation of CR protons through the DH, and of
any other CR nucleus for that matter, relies on the existence of the
propagator $\mathscr{G}_{\! p}$. This Green function translates the probability for
a CR proton injected at position $\mathbf{x}_{S} \equiv (x_{S},y_{S},z_{S})$
and time $t_{S}$ to travel through the Galactic magnetic field until it
reaches, at time $t$, an observer located at $\mathbf{x} \equiv (x,y,z)$.
The CR proton density at the Earth may be expressed as the convolution
over space and time of the Green function $\mathscr{G}_{\! p}$ with $q_{\rm acc}$
\begin{equation}
\psi(\mathbf{x},t) =
{\displaystyle \int_{- \infty}^{\, t}} dt_{S} \,
{\displaystyle \int_{\rm DH}} d^{3}\mathbf{x}_{S} \;
\mathscr{G}_{\! p} \left(
\mathbf{x} , t \, \leftarrow \, \mathbf{x}_{S} , t_{S}
\right) \;
q_{\rm acc}(\mathbf{x}_{S},t_{S}) \;\; ,
\label{psi_G_p}
\end{equation}
where $q_{\rm acc}(\mathbf{x}_{S},t_{S})$ is the CR proton injection
rate at the source located at $\mathbf{x}_{S}$ and at time $t_{S}$.
Notice that for a population $\mathscr{P}$ of point-like sources, the
injection rate $q_{\rm acc}$ is given by Eq.~(\ref{q_acc_discret})
which, once plugged in the previous expression for $\psi$, leads
to the proton flux at the Earth
\begin{equation}
\Phi =
{\displaystyle \frac{v_{p}}{4 \pi}} \times \left\{ \psi \equiv
{\displaystyle \sum_{n \in \mathscr{P}}} \;
\mathscr{G}_{\! p} \left(
\mathbf{x}_{\odot} , t = 0 \, \leftarrow \, \mathbf{x}_n , t = t_n
\right) \times q_n \right\} \;\; .
\label{psi_for_cal_P}
\end{equation}
The Green function $\mathscr{G}_{\! p}$ plays a crucial role in our analysis
and is the ideal tool to study the effect of the discreteness of the
sources. It is a solution of the CR transport equation
\begin{equation}
{\displaystyle \frac{\partial \, \mathscr{G}_{\! p}}{\partial \, t}} \, + \,
\partial_{z} \left( V_{C} \, \mathscr{G}_{\! p} \right) \, - \,
D \, \triangle \, \mathscr{G}_{\! p} \, + \,
2 \, h \, \delta(z) \, \Gamma_{p} \, \mathscr{G}_{\! p} =
\delta^{3}(\mathbf{x} - \mathbf{x}_{S}) \,
\delta(t - t_{S}) \;\; .
\label{master_equation_G_p}
\end{equation}
The resolution of this equation is presented in Appendix~\ref{appendix:green}.

\subsection{Diffusion parameters}

The computed flux depends on the diffusion parameters.
The results will be presented for three benchmark sets of diffusion
parameters, consistent with the energy dependance of the B/C ratio
\citep{2001ApJ...555..585M}. They are labelled ``min'', ``med'' and ``max'',
according to the value of $L$. The values are indicated in Table~\ref{table:parameters}.
\begin{table}[h]
\caption{Diffusion parameters for the three benchmark models}
\label{table:parameters}
\begin{tabular}{c c c c c}
\hline
model & $D_0\;(\text{kpc} \cdot \text{My}^{-1})$ & $\delta$ & $L \;(\text{kpc})$& $V_c\;(\text{km} \cdot \text{s}^{-1})$\\
\hline
min & 0.0016 & 0.85 & 1 & 13.5 \\
med & 0.0112 & 0.7 & 4 & 12 \\
max & 0.0765 & 0.46 & 15 & 5 \\
\hline
\end{tabular}
\end{table}

\section{Mean flux}

The flux at Solar position is obtained by summing the contributions of all
the point sources. The mean value can be computed, once the
distribution of sources is known. We first provide the result for a
homogeneous distribution of sources, as an analytic expression can be
found in this case. We then turn to a more realistic distribution.

\subsection{Homogeneous distribution of sources}

The average flux from one point source, drawn from a statistical
ensemble with distances ranging from $R_\text{min}$ to $R_\text{max}$
and ages from $t_\text{min}$ to $t_\text{max}$, is given by averaging
$\mathscr{G}_1\equiv \mathscr{G}_p$ over  distances and times. To get
an order of magnitude estimate, we
first consider that the sources are distributed evenly in space and
time in the Galactic disk, which is assumed to be infinitely thin and
to extend to $R_\text{max} \to \infty$. We obtain
$$\langle \mathscr{G}_1 \rangle = \frac{1}{\pi R_\text{max}^2t_\text{max}}
\int_{0}^{R_\text{max}}2\pi r \, dr \int_{t_\text{min}}^{t_\text{max}} \frac{d\tau}{4\pi DL\tau} e^{-r^2/4D\tau}  \sum_n e^{-k_n^2 D\tau}$$
Considering $\nu$ sources per unit time and unit area in the disk,
the average flux from $N = \nu
\pi R_\text{max}^2 t_\text{max}$ sources is given by
$$\langle \mathscr{G}_N \rangle = N \langle \mathscr{G}_1 \rangle=\frac{\nu}{4\pi DL}
\int_{0}^{R_\text{max}} 2\pi r \, dr\int_{0}^{t_\text{max}} \frac{d\tau}{\tau} e^{-r^2/4D\tau}  \sum_n e^{-k_n^2 D\tau}$$
By integrating over time ($t_\text{max}\to \infty$) and using (\citet{2007tisp.book.....G} 3.471.9) 
\begin{equation}
{\displaystyle \int_{0}^{\infty}} \, x^{\nu-1} \,
e^{-\beta/x-\gamma x} \; dx =
2 \left( \frac{\beta}{\gamma} \right)^{\nu/2}
K_{\nu} \left( 2 \sqrt{\beta \gamma} \right) \;\; ,
\label{eq:grad1}
\end{equation}
where $K_{\nu}$ is the $\nu$-th order modified Bessel function of the second
kind, leads to
$$\langle \mathscr{G}_N \rangle =\nu \int_0^{R_\text{max}} 2\pi r
\mathscr{G}_\text{steady}(r) \, dr$$
where $\mathscr{G}_\text{steady}$ stands for the one source steady-state propagator, given by
\begin{equation}
\mathscr{G}_{\text{steady}}
\left( \mathbf{x} \, \leftarrow \, \mathbf{x}_{S} \right)
 = \frac{1}{2 \pi D L} \sum_{n=1}^{\infty}
K_{0} \left( \rho \sqrt{\alpha_{n}/D}  \right).
\label{def:Gp_steady}
\end{equation}
Finally, using $\int x \, K_0(x) \, dx = -x\, K_1(x)$, we have
\begin{equation}\begin{split}\langle \mathscr{G}_N \rangle = \frac{\nu}{D} \sum_{n=0}^\infty \frac{2}{(2n+1)\pi} 
&\left\{ R_\text{min} K_1\left(\frac{2n+1}{2} \frac{\pi
  R_\text{min}}{L} \right)\right.\\
&\left. -
  R_\text{max} K_1\left(\frac{2n+1}{2} \frac{\pi R_\text{max}}{L} \right) \right\}\end{split}\end{equation}
For an infinite disk, $R_\text{min}=0$ and $R_\text{max} \to \infty$,
so that, using $xK_1(x) \to 1$ when $x \to 0$,
$$\langle \mathscr{G}_N \rangle = \frac{\nu L}{D} \sum_{n=0}^\infty
\left(\frac{2}{(2n+1)\pi}  \right)^2$$
It can be shown that $\sum_n 1/(n+1/2)^2 = \pi^2/2$, and finally for
the infinite disk,
$$\langle \mathscr{G}_N \rangle = \frac{\nu L}{2D}$$
which is what is also obtained by directly solving the diffusion
equation in steady state. The mean value of the flux from
randomly distributed point sources is equal to the steady-state flux
obtained with a continuous source distribution. This is also true in
more general cases, thick disk, finite radius, with wind and spallation.

\subsection{Realistic distribution of sources}
\label{sec:RealDist}

We now consider the case of a general distribution of sources, with a
radial distribution in the Galactic disk, as well as a distribution across the
thickness of the disk. 
It is quite well accepted that up to energies corresponding to the
knee, cosmic rays accelerators are supernova remnants (SNR). Unfortunately,
supernovae are pretty rare events and their spatial distribution
is difficult to measure accurately. However, as pulsars are created in
SNR and are easier to detect, it is a fair asumption that the cosmic
ray sources follow the pulsar distribution.
In this work we suppose that the radial profile of sources follows the
pulsar distribution given in \citet{2004A&A...422..545Y},
$$f_r(r_s)=\left(\frac{r_s+0.55}{8.5+0.55}\right)^{1.64} \exp\left\{
-4.01 \left(\frac{r_s-8.5}{8.5+0.55}\right)\right\} $$
where distances are expressed in kpc. The distribution along the $z$
axis is given by
$$f_s(z_s)= \exp{\left(-\frac{|z_s|}{z_0}\right)}$$ where $z_0$ is set
to the half-thickness of the Galactic disk, $z_0 = h$.
In this section, $z_s$, $r_s$, $\theta_s$ and
$t_s$ refer to the position and age of sources. The solar system is
located at $r_\odot = 8.5\;\text{kpc}$, $\theta = 0$ and $z=0$.
Moreover we define $\rho(r_s,\theta_s)$ as the distance from a source
to the Solar System: 
$$\rho(r_s,\theta_s)=\sqrt{(r_\odot -r_s \cos \theta_s)^2+r_s^2
  \sin^2 \theta_s}$$
We evaluate the scatter of the flux that arises when we consider
point-like sources with positions and ages following a given
probability distribution. The sources are supposed to be independent,
so the mean value of the propagator coming from all the $N$ sources is
just given by :
\begin{equation}\begin{split}\langle \mathscr{G}_N \rangle = &N \times \langle \mathscr{G}_1
\rangle = N A\int 2\pi \, r_s dr_s \int d\theta_s \int dz_s \int dt_s \\
& f_\theta (\theta) \, f_r (r_s) \, f_z(z_s) \, f_t(t_s) \mathscr{G}_N\left(r_s,\theta_s,t_s\right)\end{split}\end{equation}
Assuming cylindrical symmetry and a uniform age distribution, up to a
maximum age $T_\text{max}$, 
the normalization factor $A$ is given by, 
$$\frac{1}{A}=2\pi \, T_\text{max} \int_{Galaxy} r  f_r (r) \, f_z(z) dr dz$$
We find (the antisymmetric part of the propagator vanishes at $z=0$)
\begin{equation}\begin{split}
\langle \mathscr{G}_N \rangle &= N \int r_s dr_s dz_s dt_s d\theta_s A f_r(r_s) \exp\left( -\frac{|z_s|}{z_0}\right) \\
& \times \frac{1}{4 \pi D t_s} \exp\left( \frac{-V_c |z_s|}{2 D}\right) \exp( -\frac{\rho^2(r_s,\theta_s) }{4 D t_s} )\\
 & \times \sum_{n = 1}^\infty \frac{\exp\left( - (\alpha_n) t_s\right)}{C_n} \sin(k_n L)  \sin(k_n (L-|z_s|) ) \nonumber\end{split}
\end{equation}
We separate the integration and define
\begin{equation}
\langle J_z \rangle_n=\frac{\sin(k_nL)}{C_n} \int_{-L}^{L}\exp\left(-\frac{|z_s|}{z_0}\right)\exp\left(-\frac{V_c|z_s|}{2 D}\right)\sin(k_n(L-|z_s|)) 
\end{equation}
The integration is easy using the exponential form of sinus and leads to
\begin{equation}
\begin{split}
\langle J_z \rangle_n & =\frac{\sin(k_nL)}{C_n}\frac{2}{(V_c/2D +
  1/h)^2 + k_n^2}\\
& \times \left( \left(  \frac{V_c}{2D}+\frac{1}{h}\right)\sin(k_n L) -
kn \left(\cos(k_n L)- e^{-L\left(
  \frac{V_c}{2D}+\frac{1}{h}\right)}\right)\right)  
\end{split}
\end{equation}
The integration over time ($t_\text{max}\to \infty$) is made using Eq.~\ref{eq:grad1} : 
\begin{equation}
\begin{split}
\int_0^\infty \frac{1}{4 \pi D t_s} &\exp \left(-\frac{\rho^2(r_s,\theta_s)}{4 D t_s} \right) 
\exp(- \alpha_n t_s) dt_s \\ &= \frac{1}{2 D \pi}
K_0\left(\rho(r_s,\theta_s) \sqrt{\frac{\alpha_n}{D}}\right) \end{split}
\end{equation}
Then we can write the mean value as :
\begin{equation}
\langle \mathscr{G}_N \rangle = N \frac{A}{2 D \pi}
\int dr_s d\theta_s r_s f_r(r_s) f_\theta(\theta_s) \sum_n K_0 \left(
\rho(r_s,t_s)\sqrt{\frac{\alpha_n}{D}}\right) \langle J_z \rangle_n
\end{equation}
This result is integrated numerically over the two
coordinates $r_s$ and $\theta_s$. The mean flux is then given by
$$\langle \Phi \rangle = \langle \mathscr{G}_N \rangle \frac{\rm v}{4 \pi} g(E)$$
Where $g(E)$ is the source energy injection term. We checked that this calculation
gives the same result as the steady state model. 
In the next section, we perfom the calculation of the variance
associated to the flux.

\section{Statistical analysis : variance of the flux}
\label{sec:variance}

The flux obtained in the myriad model depends on the exact positions
and ages of all the sources. It is bound to be different from the mean
values computed above, at some level. How different from the mean value are likely to
be the typical fluxes in the myriad model? In order to address this
question, the usual method  is
to compute the standard deviation associated to the flux, considered
as a random variable.

\subsection{Realistic distribution of sources}
\label{sec:Variance}
The variance of the propagator for one source is defined as
$$\sigma_1^2 = \langle \mathscr{G}_1^2 \rangle - \langle \mathscr{G}_1 \rangle^2$$
The variance for $N$ independent sources is given by
$$\sigma_N^2=N \sigma_1^2$$
We must compute the average value of $\mathscr{G}_1^2$, given by
\begin{equation}
\begin{split}
\langle \mathscr{G}_1^2 \rangle  &= \int  dr_s dz_s dt_s d\theta_s A r_s f_r(r_s) \exp\left( -\frac{|z_s|}{z_0}\right) \\
& \times \left( \frac{1}{4 \pi D t_s}\right)^2 \exp\left( -\frac{V_c |z_s|}{D}\right) \exp\left( -\frac{\rho^2(r_s,\theta_s) }{2 D t_s} \right)\\
& \times \sum_{n = 1}^\infty \sum_{m = 1}^\infty \frac{\exp\left( -
  (\alpha_n + \alpha_m) t_s\right)}{C_n C_m} \\
& \sin(k_n L ) 
\sin(k_m L) \sin(k_n (L-|z_s|) ) \sin(k_m(L-|z_s|))
\end{split}
\end{equation}
At fixed $m$ and $n$, we define
\begin{equation}
\begin{split}
\langle \mathscr{G}_z^2 \rangle_{n,m} \equiv \frac{ \sin(k_n L)
  \sin(k_m L) }{C_n C_m} \int_{-L}^L & dz_s \exp \left( -|z_s| \left(
\frac{V_c}{D} + \frac{1}{h} \right )\right) \\
&\sin(k_n (L - |z_s|)) \sin(k_m (L - |z_s|))
\end{split}
\end{equation}
This is easily computed, and we can integrate the time distribution to
get the variance from sources of all ages : 
\begin{equation}
\begin{split}
\langle \mathscr{G}_1^2 \rangle = A \sum_{n,m} \langle \mathscr{G}_z^2
\rangle_{n,m} \int & dt_s dr_s d\theta_s r_s f_r(r_s)
\exp(-(\alpha_n+\alpha_m) t_s) \\
& \times \frac{1}{16 \pi^2 D^2 t_s^2} \exp\left(-\frac{2 \rho^2(r_s,\theta_s)}{4 D t_s}\right)
\end{split}
\end{equation}
Using Eq.~\ref{eq:grad1} and considering $t_\text{max}\to \infty$
\begin{equation}
\begin{split}
 \int_0^\infty &dt_s \left( \frac{1}{4 \pi D t_s} \right)^2 \exp \left(
 - \frac{\rho^2}{2 D t_s}\right) \exp (- (\alpha_n +\alpha_m)t_s)\\
&=
 \sqrt{2 (\alpha_n +\alpha_m) D}\frac{1}{8 \pi^2
   D^2}\frac{1}{\rho}K_1\left( \rho \sqrt{\frac{ 2(\alpha_n
     +\alpha_m)}{D}}\right)
\end{split}
\end{equation}
we find
\begin{equation}
\begin{split}
\langle \mathscr{G}_1^2 \rangle  = &A \sum_{n,m}  \langle \mathscr{G}_z^2
\rangle_{n,m} \frac {1}{8 \pi^2 D^2} \sqrt{2 \left( \alpha_m+\alpha_n
  \right) D} \int_{0}^{R_\text{max}} r_s dr_s \\ & \times \int_{0}^{2\pi} d \theta_s \frac{1}{\rho(r_s,\theta_s)} K_1\left(\rho(r_s,\theta_s)\sqrt{2 \frac{\alpha_m+\alpha_n}{D}}\right) f_r(r_s)
\end{split}
\end{equation}
We are interested in the behaviour of this quantity at the lower bound
in $r$. When $r \sqrt{2 (\alpha_m+\alpha_n)/D} \ll 1 $, the
property $K_1(x) \to 1/x$ as $x \to 0$ can be used to show that the above
integral diverges as $\ln (R_\text{min})$ as the lower bound
$R_\text{min}$ of the spatial integral goes to zero. The spatial
distribution probability is $f_r \sim 1$ when $r \to 0$, so that
\begin{equation}
\langle \mathscr{G}_1^2 \rangle  \sim \frac {A}{4 \pi D} \ \sum_{n,m}  \langle \mathscr{G}_z^2
\rangle_{n,m} \ln R_\text{min}
\end{equation}

\subsection{Divergence of the variance: how bad is it?}

The mean value, computed considering $r_s$ from $r_s = 0$ to $r_s = 20$ kpc
and $t_s$ from $t_s = 0$ to $t_s = \infty$ is equal to the flux obtained in
the steady-state model, as commonly obtained in propagation theories. But
as we have just seen, the calculation related to the standard deviation of
this quantity diverges, when we allow the possibility of sources that are
close ($r_s \to 0$) and young ($t_s \to 0$), as already noticed before
(\citep{2011arXiv1105.4521B}).

The standard deviation is usually interpreted as the typical spread of
the random values around the mean, and a large standard deviation could be
interpreted as if the actual value of the flux had a disturbingly large
probability to be very far from the mean value.

\begin{figure*}[h!]
\centerline{
\includegraphics[width=7cm]{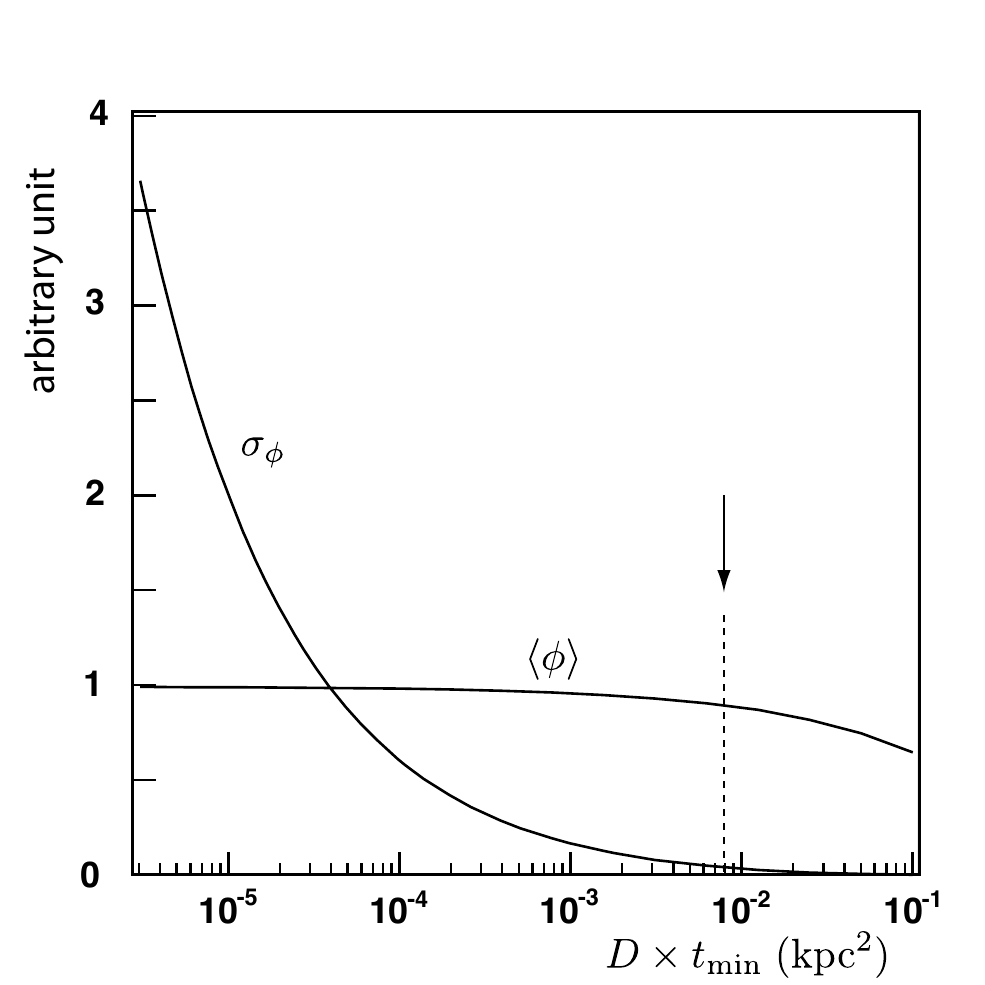}}
\caption{Evolution of the mean value and of the standard deviation of
the flux with the age cut-off $t_\text{min}$. The cut-off obtained
with the prescription $t_\text{min} = R_\text{max}/\sqrt{4\nu D(E)}$
(see text) is indicated with an arrow. This is for the max model, at
an energy of 10 TeV.}
\label{fig:variancecutoff}
\end{figure*}

One could argue that the problem we considered is physically
irrelevant, as we know for sure that there is no supernova remnant
with zero age and null distance to the Earth. One can impose a lower
cut-off in ages and distances, based on observations. However, even
with reasonable values for the cut-off, the variance can be finite but
large (see section \ref{subsec:cutoff}).
One can also adjust the cut-off in order to eliminate the very rare
events that make the standard deviation very large, without contributing
significantly to the the mean value. This is the approach adopted by
\citet{2011arXiv1105.4521B}. These authors use a cut-off given by
$t_\text{min} = R_\text{max}/\sqrt{4\nu D(E)}$. This condition gives
a fair order of magnitude of the spread of the values around the mean.
It is difficult though to interpret it in rigourous statistical terms.
Actually, the value of the variance depends quite strongly on the exact
position of the cut-off, as featured in Fig.~\ref{fig:variancecutoff}
where the evolution of the mean and the standard deviation of the total
proton flux derived with the max model at 10 TeV is presented. Choosing
$t_\text{min}/2$ or $2 \times t_\text{min}$ rather than $t_\text{min}$,
for instance, has a small effect on the mean but has a drastic effect
on the standard deviation. The value chosen by \citet{2011arXiv1105.4521B}
is indicated by an arrow on the figure. The mean obtained with this cut-off
is about 10 \% smaller than the true mean. A lower cut-off would give
a more precise mean value, but a much larger variance.

This confusing situation, where some rare events have a very small
contribution to the mean, but give rise to a very large standard deviation,
is not uncommon in physics (Levy flights, Cauchy distribution).
The total flux $\phi$ is the sum of a myriad of $N$ individual contributions
$\varphi_{i}$ from single sources. Assuming that these contributions
are not correlated with each other allows to relate simply the means
and variances of the total and single source fluxes through
\begin{equation}
\langle \phi \rangle = N \times \langle \varphi \rangle
\;\;\; {\rm and} \;\;\;
\sigma_{\phi}^{2} = N \times \sigma_{\varphi}^{2} =
N \langle \varphi^{2} \rangle \, - \, \frac{\langle \phi \rangle^{2}}{N}.
\end{equation}
We show in the appendix that the probability $p(\varphi)$ of measuring
a single source flux $\varphi$ is given by
$p(\varphi) \propto \varphi^{-8/3}$ for objects located in a thick disk and
$p(\varphi) \propto \varphi^{-7/3}$ for sources located in a thin disk, in
the limit where $\varphi \to \infty$. The standard deviation $\sigma_{\phi}$
of the total flux is related to the integral
$\langle \varphi^{2} \rangle = \int \varphi^2 \, p(\varphi) \, d\varphi$,
which diverges as $\varphi \to \infty$. For such a distribution with an
infinite second moment, the central limit theorem does not hold, at least
in its usual form. In this case, the standard deviation is not a good
estimator of the typical spread of the values around the mean (in
particular the probability distribution function $p(\phi)$ is no longer
Gaussian), and there is no use trying to regularize this quantity by
applying cut-offs.
Notice that the average value $\langle \phi \rangle$ is still well-defined,
as $\langle \varphi \rangle = \int \varphi \, p(\varphi) \, d\phi$ is
convergent, and more to the point, the confidence intervals of the global
probability distribution $p(\phi)$ are well-defined too, since the integral
of $p(\varphi)$ is well-behaved. The spread of the random values of the total
flux around its mean $\langle \phi \rangle$ can then be studied by computing
the quantiles associated to the probability distribution $p(\phi)$, rather
than using the standard deviation which has no clear statistical relevance
in this case.

All this discussion arises from the fact that, in the myriad model,
there is a non-vanishing probability to find sources which are arbitrarily
close and young. Although the variance $\sigma_{\phi}$ cannot be defined
in this case, we can still infer the global distribution function $p(\phi)$
and delineate intervals inside which the total fux $\phi$ is mostly expected.
Another possibility is to take advantage of the solid information which has
been collected on these young and nearby sources, whose statistical properties
are disconcerting. We can actually remove a space-time region around the Earth
containing the very close and very young sources of our Galaxy, and replace
them by a catalog, built from observations.

In what follows, we investigate these points in further details. First we study
how the standard deviation $\sigma_{\phi}$ can be regularized by applying some
cut-offs related to astronomical information about the sources. We then focus
on the confidence intervals of $p(\phi)$ and we show that they are well-defined,
with no need of a particular cut-off. Finally, we take into account the
astronomical knowledge about the local sources of cosmic rays and explore in
greater detail how introducing such a catalog improves our statistical analysis.

\subsection{Regularisation by cutting off the ages}
\label{subsec:cutoff}

From observations of the Solar neighbourhood, we have a good idea of the
distribution of sources which are young and close. Given the age $t_\text{min}$
of the youngest local supernova remnant, one can compute the mean value and
standard deviation of the total flux $\phi$, by applying that cut-off $t_\text{min}$
to the age distribution.

\begin{figure*}
\centering
\includegraphics[width=0.5\textwidth]{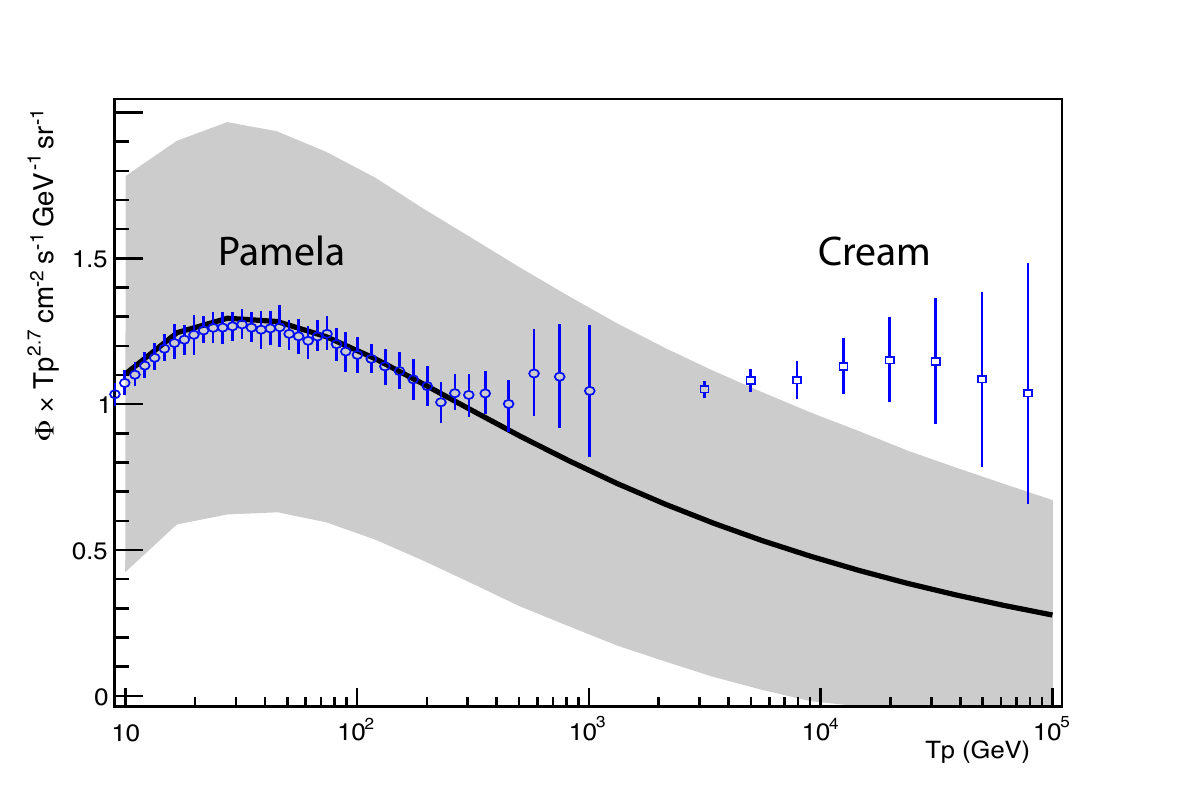}
\caption{
Proton flux obtained for the min diffusion model (see below) and a source
frequency $\mu = 3 \; \text{century}^{-1}$. The band displays the standard
deviation $\sigma_{\phi}$ obtained analytically, with a lower cut-off on the
ages of the sources $t_\text{min} = 100 \; \text{years}$. The blue symbols
indicate the flux measured by
CREAM \citep{2011ApJ...728..122Y} and
PAMELA \citep{2011Sci...332...69A}.
}
\label{fig:varianceStat}
\end{figure*}

For the sake of illustration, Fig.~\ref{fig:varianceStat} features
the mean value $\langle \phi \rangle$ and the standard deviation
$\sigma_{\phi}$ of the total proton flux using a lower cut-off of
$t_\text{min} = 100 \; \text{yr}$ in the source distribution, for the
min set of propagation parameters. On the same figure we have also plotted 
the data points from the CREAM and PAMELA experiments. With the chosen value
for the cut-off, the standard deviation at high energies remains of the same
order of magnitude as the flux. The relative value of the standard deviation,
${\sigma_{\phi}}/{\langle \phi \rangle}$, is fairly independent of the energy.
This trend can be shown analytically to hold for sources located in a thin disk.
In the framework of a purely diffusive model (no wind/spallation), the relative
dispersion can be actually approximated by
\begin{equation}
\frac{\sigma_{\phi}}{\langle \phi \rangle} \sim
\frac{R}{4 L \sqrt{2 \nu t_\text{min}}} \;\; .
\end{equation}
This ratio does not depend on the diffusion coefficient $D$,
hence it does not depend on energy. It is of order or unity for
$t_\text{min} \sim 100 \; \text{yr}$.

For sources distributed in a disk having a finite thickness $h$, one
gets the same result as before as long as $h \ll \sqrt{D t_\text{min}}$.
In the opposite limit, one finds
\begin{equation}
\frac{\sigma_{\phi}}{\langle \phi \rangle} \propto D^{1/4} \;\; .
\end{equation}
In both cases, the relative standard deviation does not vary much with
energy.

The value of $\sigma_{\phi}/\langle \phi \rangle$ depends sensitively on
the chosen cut-off $t_\text{min}$. For a very low value of $t_\text{min}$,
the standard deviation is very large. Conversely, for larger values of
$t_\text{min}$, the standard deviation decreases, and the average value
$\langle \phi \rangle$ starts to be affected by the cut-off, as shown in
Fig.~\ref{fig:variancecutoff}.

\subsection{The variance is infinite but the confidence levels are finite}
\label{sec:DivVar}

The variance of the flux being infinite does not necessarily imply
that the random values are typically very different from the mean.
To illustrate this affirmation, consider the case of a unique point-like
steady-state source located in the Galactic disk, with cosmic ray diffusion
taking place in a boundless space. The solution of the diffusion equation
is given by $\varphi = a/r$ where $a$ is a constant.
Assuming that this source is uniformly distributed inside a disk of radius
$R$ leads to the probability distribution function
\begin{equation}
dp = \frac{2 \pi r \, dr}{\pi R^2} = \frac{2 r\, dr}{R^2}\;\; .
\end{equation}
We can readily infer the mean flux
\begin{equation}
\langle \varphi \rangle = \int_0^R \, \frac{a}{r} \, \frac{2r\, dr}{R^2}
= \frac{2a}{R} \;\; ,
\end{equation}
and the average value of the flux squared
\begin{equation}
\langle \varphi^2 \rangle = \int_\epsilon^R \, \frac{a^2}{r^2} \, \frac{2r\, dr}{R^2}
= \frac{a^2}{R^2} \ln \left( \frac{R}{\epsilon} \right) \;\; ,
\end{equation}
where we have introduced a cut-off value $\epsilon$ at the lower end of the
radial distribution, to exhibit the divergence of $\langle \varphi^2 \rangle$.
The variance of $\varphi$ goes to infinity as $\epsilon \to 0$.

However, the distribution of $\varphi$ (which is what we are really
interested in) is well-behaved. From the relation between $r$ and
$\varphi$, we can write
$dr = {a \, d\varphi}/{\varphi^{2}}$ so that
\begin{equation}
dp(\varphi) = \frac{2 r \, dr}{R^{2}} =
\frac{2 a^{2} \, d\varphi}{R^{2} \, \varphi^{3}} \;\; .
\end{equation}
The probability that the flux is smaller than a given value $\Phi$ may be
expressed as
\begin{equation}
P(<\Phi) = \int_{\varphi(r)}^{\varphi(R)} dp(\varphi)
= 1 \, - \, \frac{a^{2}}{R^{2} \Phi^{2}} \;\; ,
\end{equation}
provided that $\Phi > \Phi_0 \equiv a/R$. Introducing the Heavyside
distribution $\Theta$ leads to
\begin{equation}
P(>\Phi) = \frac{a^{2}}{R^{2} \Phi^{2}} \,
\Theta \left( \Phi - \frac{a}{R} \right) =
\frac{\langle \varphi \rangle^{2}}{4 \, \Phi^{2}} \,
\Theta \left( \Phi - \frac{\langle \varphi \rangle}{2} \right) \;\; .
\end{equation}
The probability that $\Phi > 10 \, \langle \varphi \rangle$ is only 1/400,
even though the variance is infinite. Actually, the flux is more likely to
be lower than the mean value, whereas one might have guessed the opposite,
considering the divergence of the variance.

When $N$ sources are considered, the mean flux value and the variance
are both just multiplied by $N$. The probability distribution $p_N(\phi)$
for the flux can be obtained by recurrence from
\begin{equation}
p_N(\phi) = \int p(\varphi) \, p_{N-1}(\phi - \varphi) \, d\varphi \;\; .
\end{equation}
These are displayed in Fig.~\ref{fig:tirages}. The variance still diverges.
In the high-$\phi$ region, the flux is dominated by the contribution of a
single source and the probability distribution is given by
\begin{equation}
\frac{dp_N}{d\phi} = N \, \frac{dp}{d\varphi} =
\frac{2 N a^{2}}{R^{2} \phi^{3}} \;\; .
\end{equation}
For $N=100$ sources, the probability that $\phi > 2 \, \langle \phi \rangle$
is $2.5 \times 10^{-3}$ and $P(\phi > 10 \, \langle \phi \rangle)$ is vanishingly
small.

\begin{figure*}[h!]
\centerline{\includegraphics{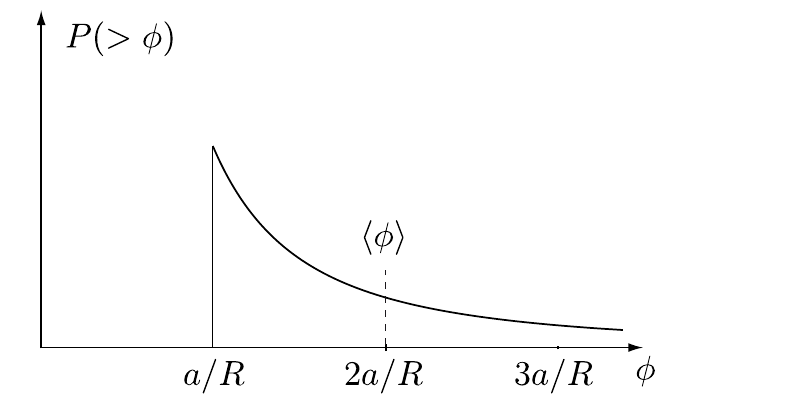}}
\caption{
Probability that the flux $\varphi = a/r$ is greater than $\phi$, for a
unique point source drawn randomly in the disk, within a distance $R$.
}
\label{fig:variance_example}
\end{figure*}
\begin{figure*}[h!]
\centerline{\includegraphics[width=8cm]{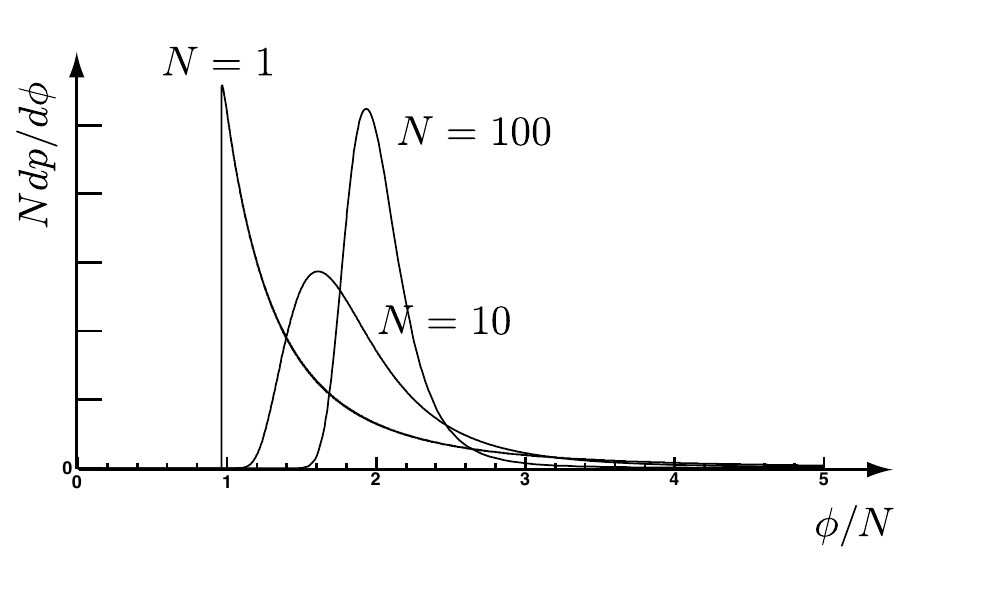}}
\caption{
Probability distribution of $\phi/N$ for $N=1$, 10 and 100 sources.
}
\label{fig:tirages}
\end{figure*}

\begin{figure*}[h!]
\centerline{
\includegraphics[width=11cm]{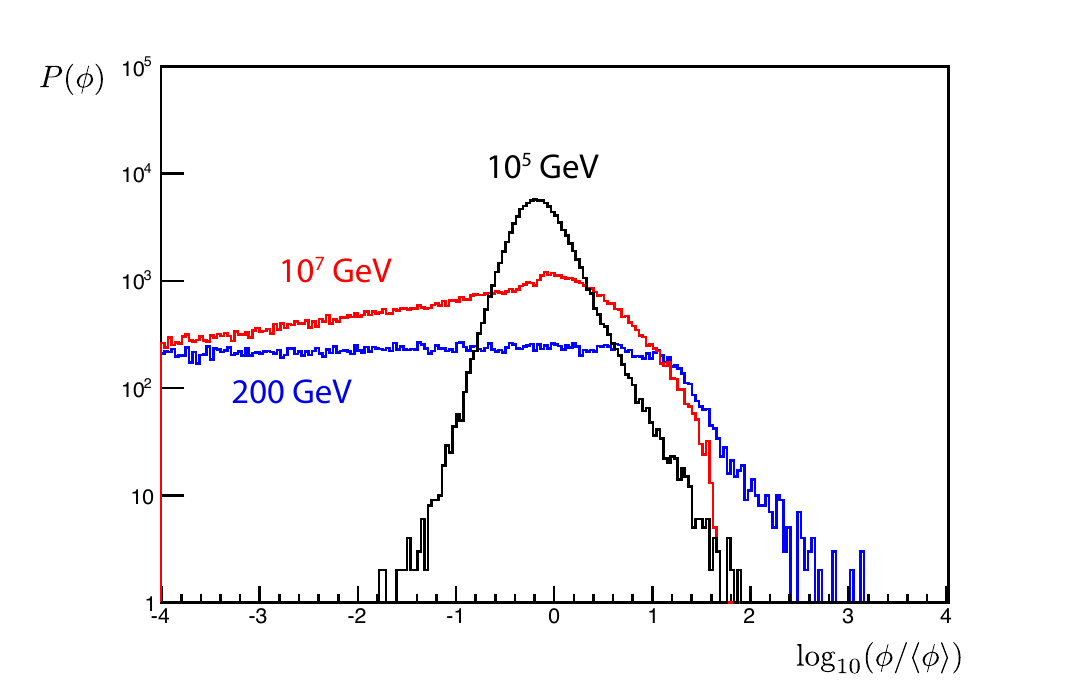}}
\caption{
Examples of histograms of fluxes for the ``min'' cosmic ray propagation model.
}
\label{fig:Histos}
\end{figure*}

If we now consider time-dependent sources spread homogeneously inside
an infinite DH with pure diffusion, the variance is given by the integral
\begin{equation}
\sigma_{\varphi}^{2} \propto \int dt \int 4 \pi r^2 \, dr \,
\frac{1}{(4 \pi D t)^{3/2}} \, e^{-r^2/2Dt} \;\; ,
\end{equation}
which diverges with the lower cut-off in ages as
${1}/{\sqrt{t_\text{min}}}$. For a 3D homogeneous distribution of steady-state
sources, $\sigma_{\varphi}$ does not diverge (see Table~\ref{table:divergence}).
For the sake of illustration, Fig.~\ref{fig:Histos} presents histograms for
the flux obtained with the propagator discussed in Sec. 2, at several energies. 

\begin{table}
\caption{Divergence of the variance}
\label{table:divergence}
\begin{tabular}{c p{3cm} p{3cm}}
\hline
& 2D &3D\\
\hline
steady-state & $\langle \varphi \rangle$ finite\newline $\langle \varphi^2
\rangle \to \infty$   &  $\langle \varphi \rangle$ finite\newline $\langle \varphi^2
\rangle$ finite  \\
time-dependent & $\langle \varphi \rangle$ finite\newline $\langle \varphi^2
\rangle\to \infty$  &  $\langle \varphi \rangle$ finite\newline $\langle \varphi^2
\rangle\to \infty$\\
\hline
\end{tabular}
\end{table}

\subsection{Quantiles of the total flux}

\begin{figure*}[h!]
\centerline{\includegraphics[width=0.5
    \textwidth]{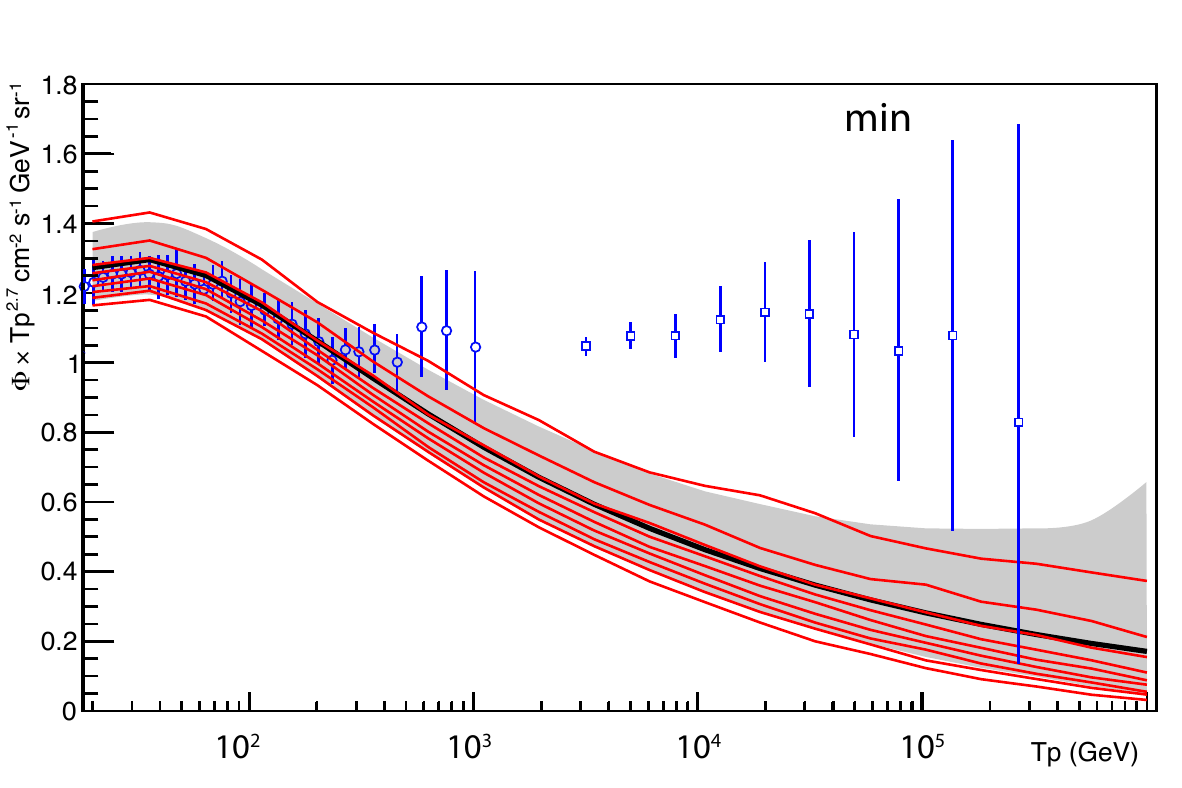}\hfill\includegraphics[width=0.5
    \textwidth]{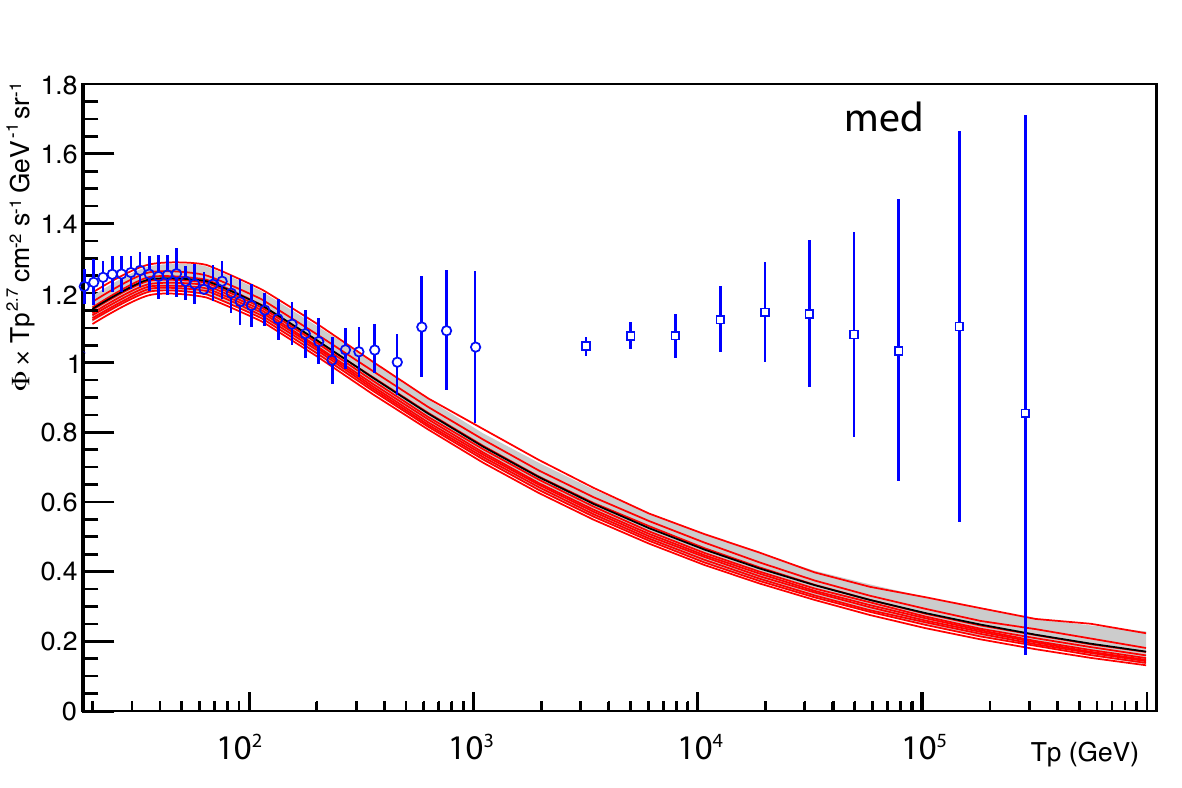}}
\centerline{\includegraphics[width=0.5\textwidth]{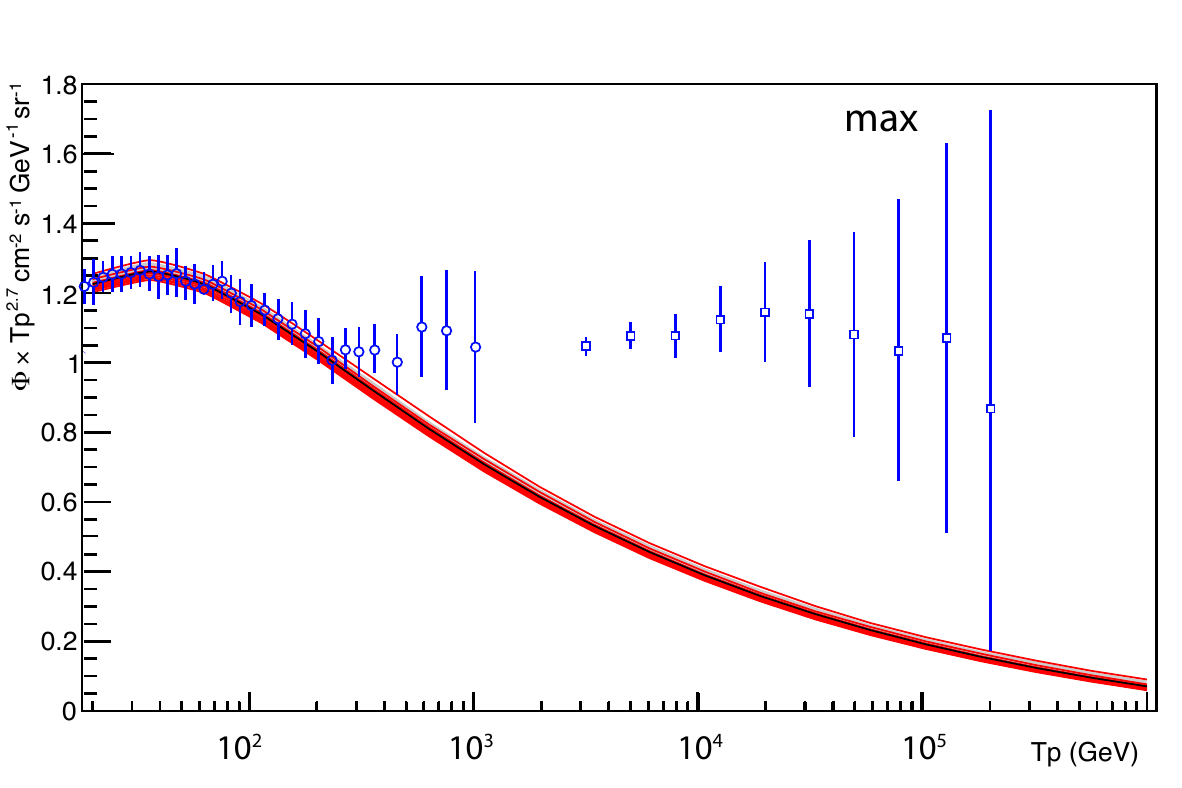}}
\caption{Confidence intervals for the flux from all the sources : the
  red curves show the 10 $\%$ quantiles, the black curves show the
  mean value and the grey band show the 68\% confidence interval for
  the flux, for $\nu=3\;\text{century}^{-1}$.}
\label{fig:confidence_total}
\end{figure*}

From now on, we show the quantiles associated to the fluxes. These
are defined as regions with a given probability to find the flux. 
In Fig.~\ref{fig:confidence_total}, we present ``deciles'' (10 \% quantiles), as well as
68 \% confidence intervals (CI), which are more familiar, as they
correspond to 1-$\sigma$ intervals for gaussian distributions. 

In order to compute these figures, we ran a Monte Carlo simulation
over $\sim$ 3000 populations.
We also adjusted the injection rate so that the mean flux follows the PAMELA data below 200
GeV :
\begin{equation}
 3.53 \times 10^{-3} \times \left( 1 - e^{-(T_p / 2.5)^{0.9}} \right)  \left(\frac{T_p}{10}\right)^{-2.5}
   \left( \frac{1+T_p}{16} \right)^{-0.42}
 \end{equation}
in $\text{cm}^{-2} \text{s}^{-1} \text{sr}^{-1} \text{GeV}^{-1}$.
 
The spread of the fluxes around the mean is not sensitive to this
choice.
In spite of its infinite variance, the flux has a well-defined
probability distribution as featured in Fig.~\ref{fig:confidence_total}.
The possible flux lies on a band whose thickness depends sensitively
on the propagation model. In particular, the width increases when the
thickness $L$ of the diffusive halo decreases.

At high energy, the CREAM and PAMELA data cannot be explained in
the med and max cases. For the min case, taking $\nu = 1\;\text{century}^{-1}$ would
enlarge the uncertainty band, so that the explanation in terms of
sources becomes more probable.

\section{Known local sources -- The catalog}

\subsection{Regularization using a catalog}

The situation described above occurs as long as we know nothing of the positions
and ages of the CR sources. The young and nearby objects are responsible for the
divergence of the flux variance and potentially lead to the problems encountered
above in the statistical analysis.
However, we actually have data concerning the distribution of nearby sources,
for which catalogs are available. A natural way then to regularize the variance
is to separate the sources in two lots.
The first set contains the young and local sources, which can be extracted from
the catalogs. The second group, on which little information is known, comprises
the old or distant sources and will be treated according to the statistical
analysis of Sec.~\ref{sec:variance}. This procedure allows to regularize the
variance of the flux in the most natural way while reducing its uncertainties.
Following~\citet{2010A&A...524A..51D}, we have used two catalogs.

\noindent
{\bf (i)} The Green survey~\citep{Green2009} compiles various informations
on supernova remnants, but fails to systematically provide their ages or the
precision with which their distances from the Earth have been determined. A
quite thorough bibliographic work has been summarised in the appendix
of~\citet{2010A&A...524A..51D}, which we have borrowed as a complement of the
Green catalog. In total, we have collected 27 local SNR with their ages,
distances from the Sun and, when possible, the corresponding observational
uncertainties.

\noindent
{\bf (ii)} Pulsars are not expected to be sources of primary CR nuclei. As
residues of supernova explosions, they are nevertheless a good tracer of old
SNR too old to be detected directly in radio waves. Moreover,
being point-like objects, their distances from the Sun is way easier to measure.
Their ages can also be estimated precisely through spin-down. After removing
millisecond pulsars from the ATNF catalog~\citep{2005AJ....129.1993M} (by
selecting $\dot{P} > 5 \times 10^{-18}$) and objects associated with known
SNR, we are left with 157 objects with their ages and distances.

\begin{figure*}
\centering
\includegraphics[width=0.4\textwidth]{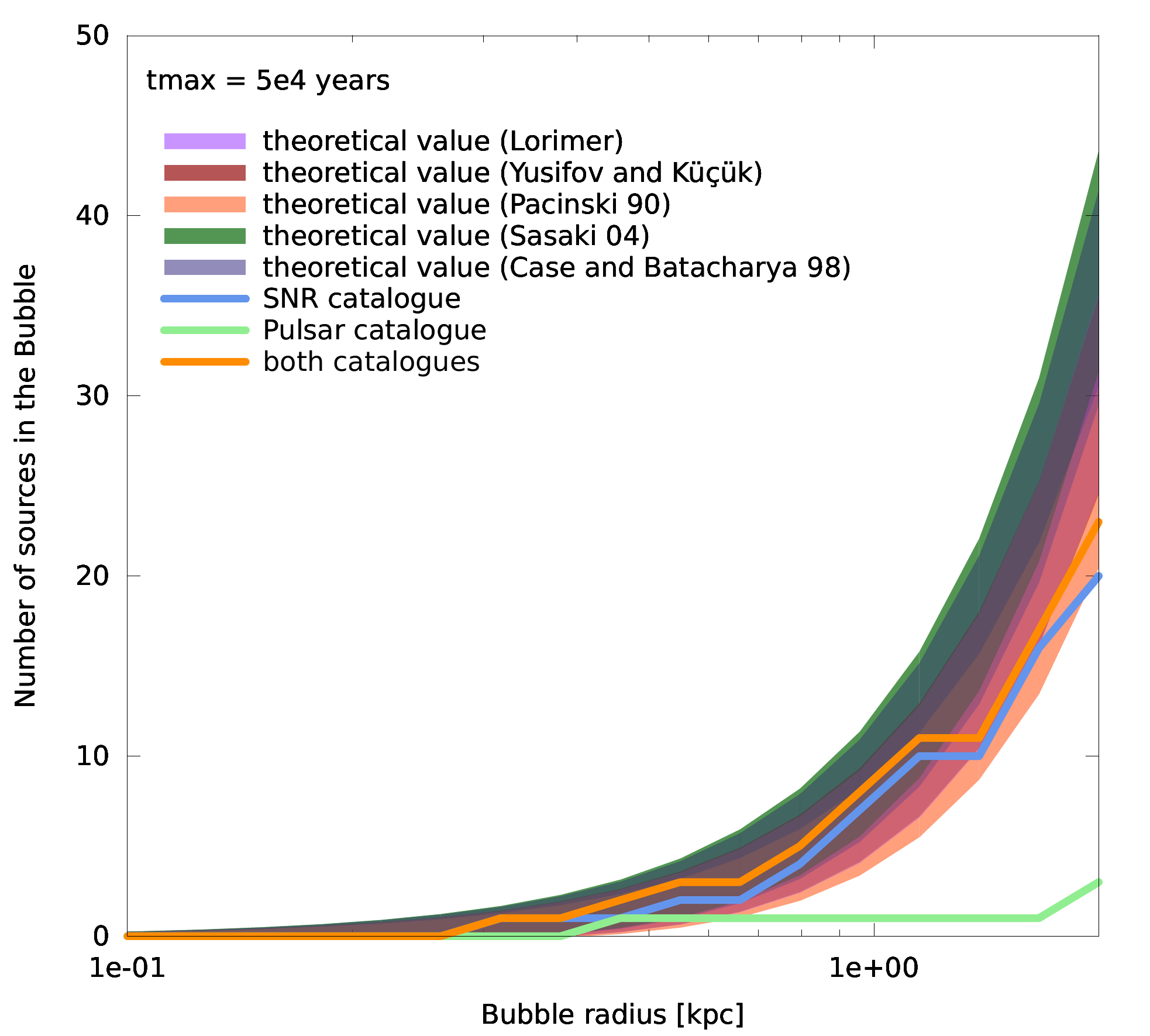}
\includegraphics[width=0.4\textwidth]{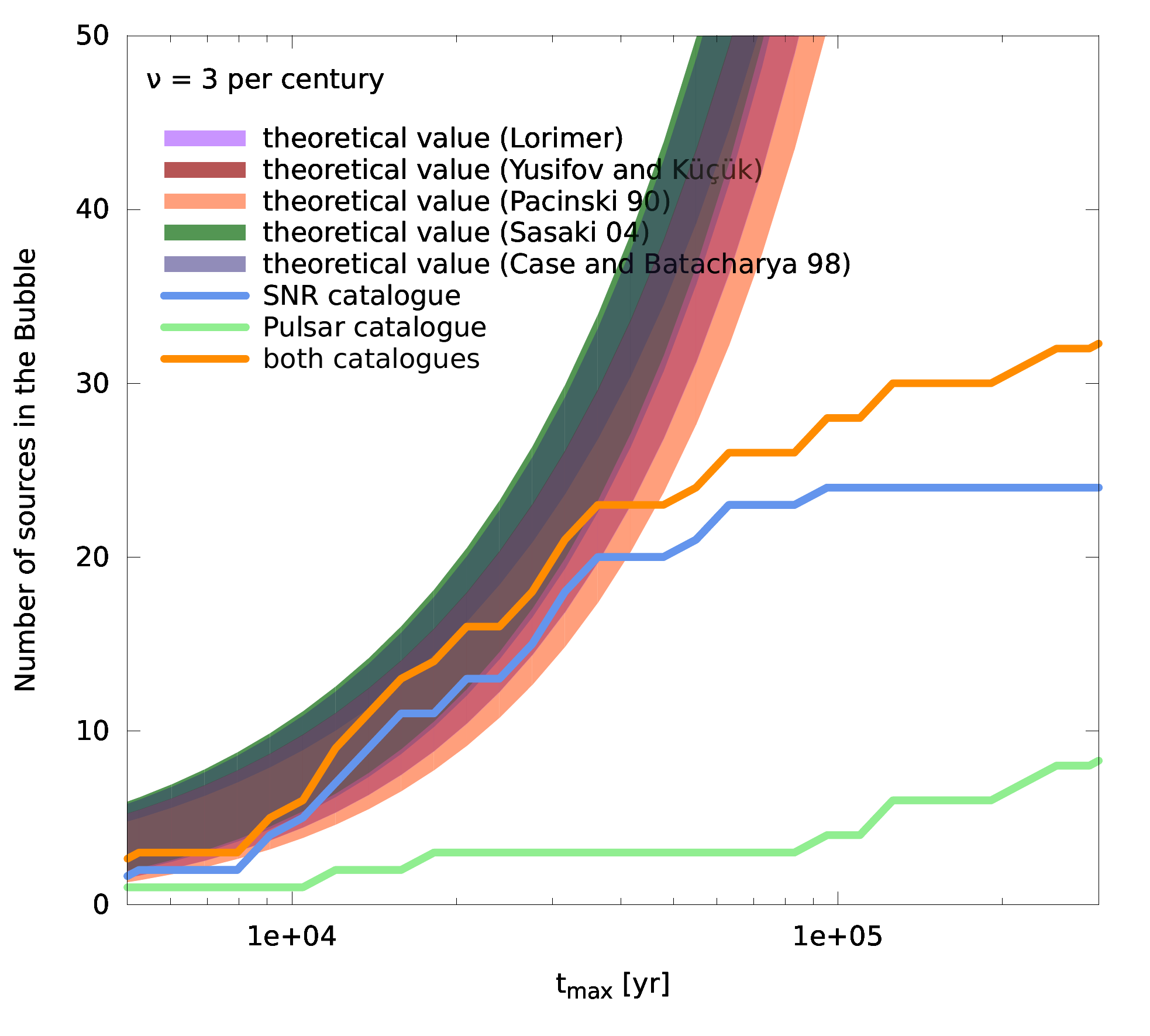}
\caption{
The bands indicate the theoretical expected number of sources within distance $r$
and for a given age (left), or with ages less than $t$ within a given distance (right),
for a supernova rate of 3 explosions per century in the Galaxy.
The width of the bands gauges the uncertainty in the number of local sources due to the
shot-noise effect. The curves feature the cumulated number of sources in our catalog.
This one appears to be complete for
$t < 5 \times 10^{4} \; \text{y}$ and $r < 2 \; \text{kpc}$.
}
\label{fig:NSources}
\end{figure*}

It is fair to ask whether these 27+157 sources are representative of the local
environment. As featured in Fig.~\ref{fig:NSources}, the number of objects
found in the catalogs is in good agreement with what can be inferred from various
Galactic distributions found in the literature, provided that the supernova
explosion rate is, on average, approximately equal to 3 per century.
At least, this is true within the 2~kpc nearby the Sun, for sources younger than
30,000 years. At later ages, SNR become too dim to be detected and
the Green survey cannot be trusted anymore. However, the catalogs are in rather
good agreement with theoretical expectations and do not suffer from major biases,
at least not more than the theoretical models.
In this work, we define the ``local'' region as the domain extending 2~kpc around
the Sun, with sources younger than 50,000 years. A rate of 1 supernova explosion
per century in our Galaxy is also quoted in the literature \citep{2010A&A...524A..51D}. If so, we would be in
a locally high-density region of sources. We have plotted the flux considering this
low rate, but will discuss the validity of this assumption in a forthcoming letter.

\begin{figure*}[h!]
\centerline{\includegraphics[width=11cm]{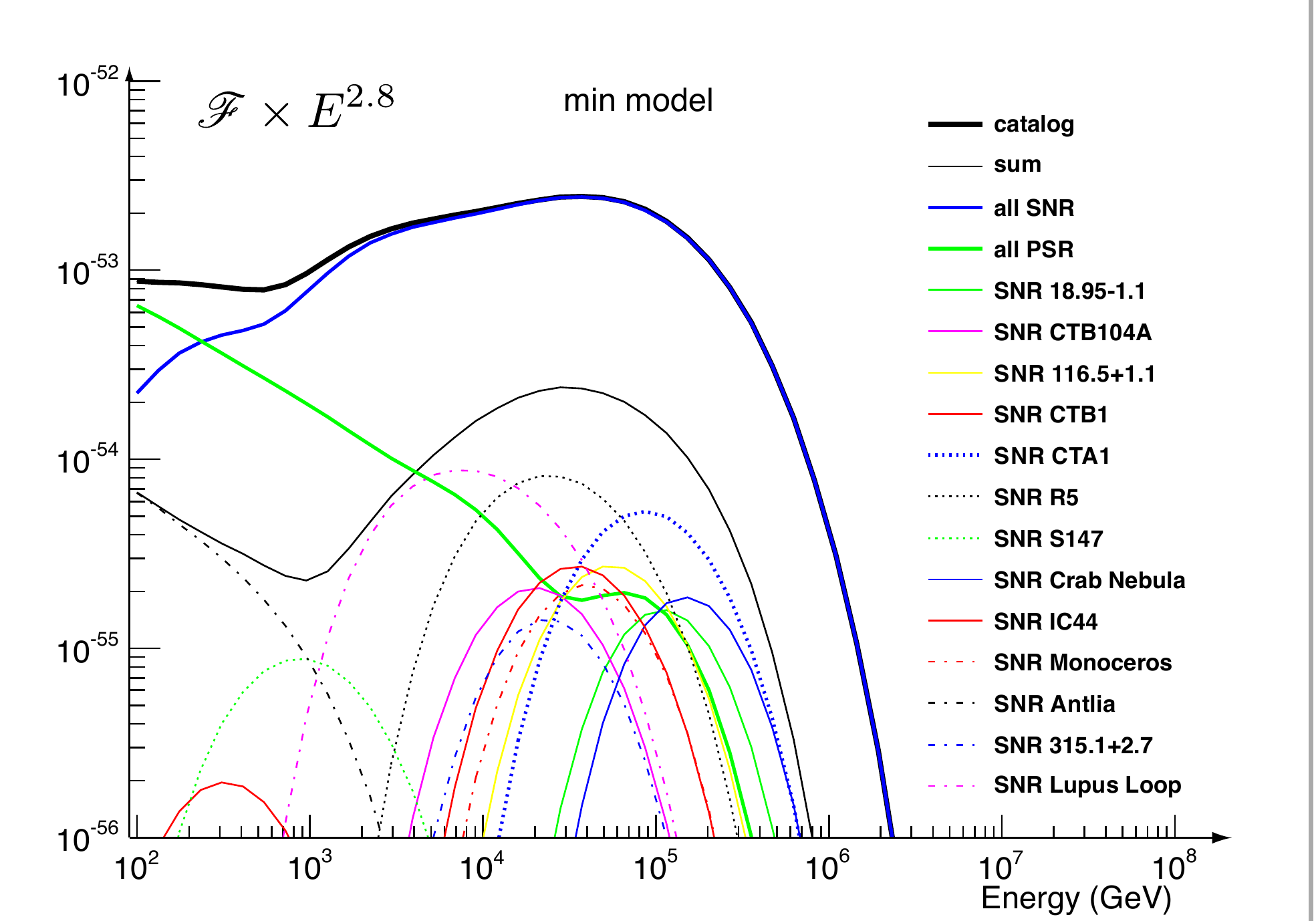}}
\caption{
The CR proton flux is plotted as a function of energy for the SNR and
pulsars which dominate over the other objects of our catalog.
}
\label{fig:SNRvsPulsar}
\end{figure*}

The CR proton flux produced by the SNR and pulsars of our catalog is presented in
Fig.~\ref{fig:SNRvsPulsar}, for the ``min'' CR propagation benchmark model. In the
PAMELA energy region, the flux is dominated by pulsars whereas SNR come into play
at the energies of CREAM.

\subsection{Is the Catalog probable?}

\begin{figure*}[h!]
\centerline{
\includegraphics[width=0.5\textwidth]{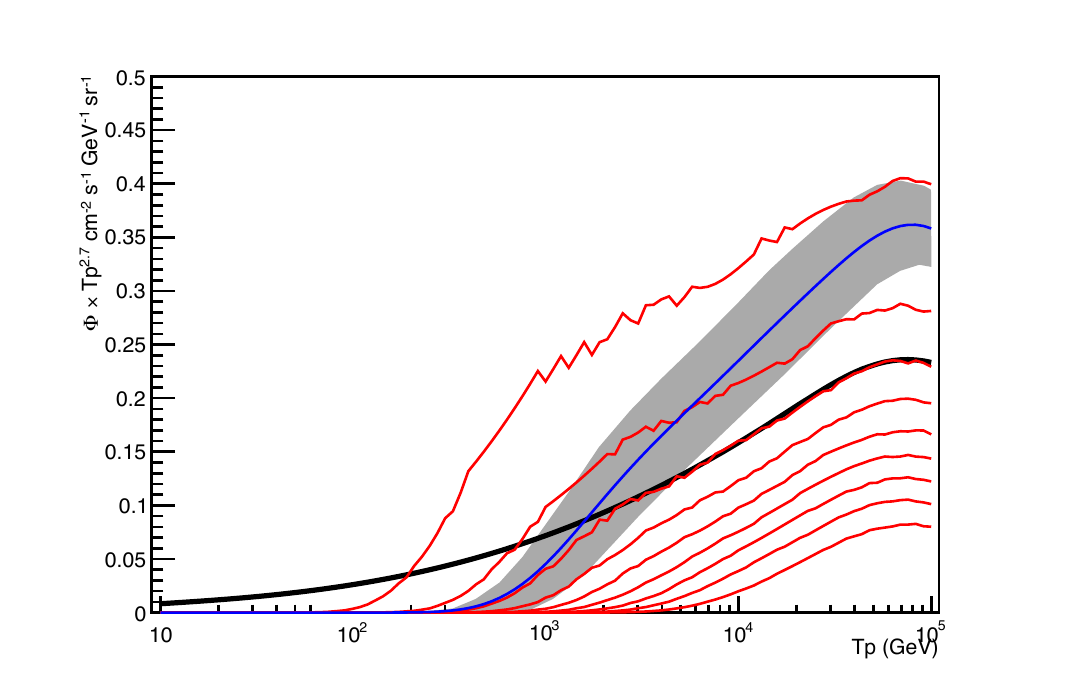}
\hfill
\includegraphics[width=0.5\textwidth]{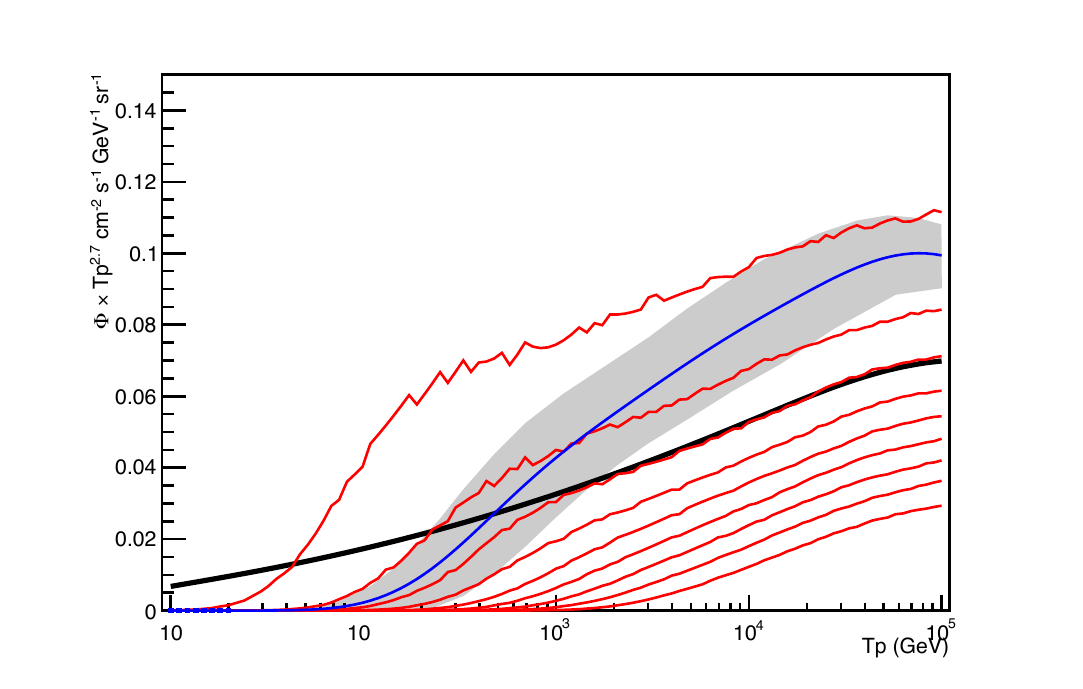}}
\centerline{
\includegraphics[width=0.5\textwidth]{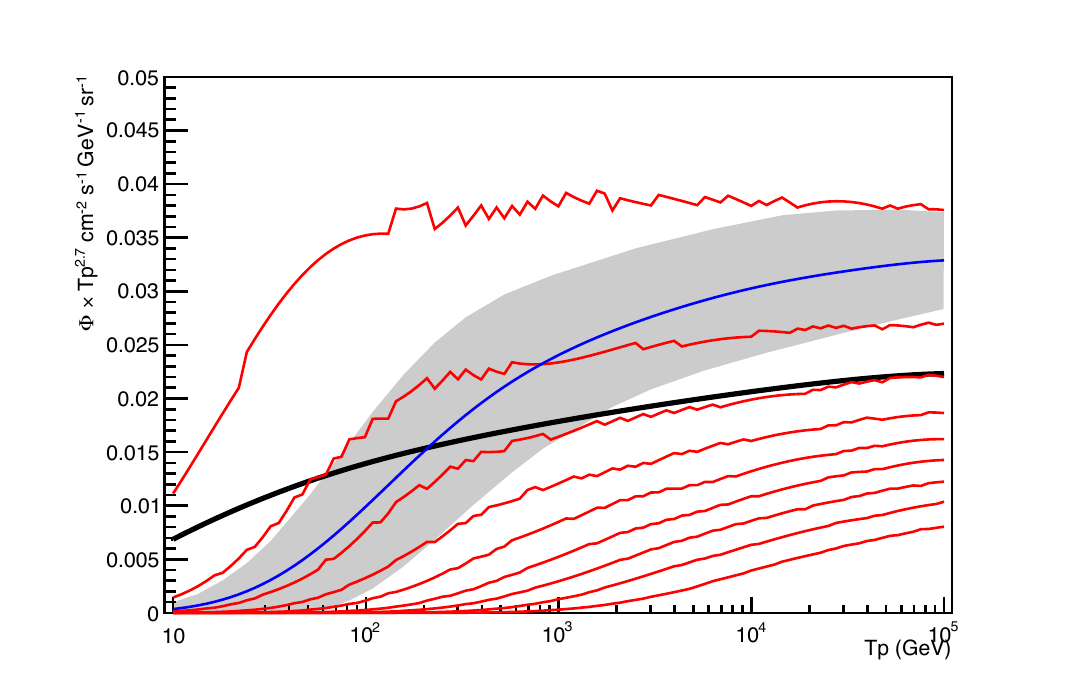}}
\caption{
In each panel, the red lines correspond to the 10$\%$ quantiles of the CR proton
flux generated by random sources drawn in the local region. The mean value is
indicated by the black curve. The flux produced by the sources belonging to our
catalog is represented in blue. The systematic errors associated to the uncertainties
on the distances and ages of these objects span the grey band (standard deviation).
All curves are derived with a supernova explosion  rate of
$\nu = 3 \; \text{century}^{-1}$.
The min, med and max propagation benchmark models are considered.
}
\label{fig:confidence_local}
\end{figure*}

In order to check the plausibility of our catalog, we compare in
Fig.~\ref{fig:confidence_local} the flux $\phi_{\rm cat}$ yielded by the sources
which it contains (blue curve) to the flux $\phi_{\rm loc}$ from a set of populations
drawn randomly inside the local region (black curve). As explained above, we did not
compute the variance of these random populations, but derived through Monte-Carlo
simulation the confidence intervals.
Assuming an explosion rate of 3 events per century in the Galaxy, we find that
the flux $\phi_{\rm cat}$ yielded by the objects of our catalog lies within
the 1-$\sigma$ confidence interval surrounding the mean value $\phi_{\rm loc}$.
This is true for the min, med and max benchmark sets of CR propagation parameters.
We conclude that the theoretical source distribution and explosion rate, which we
have chosen here, are in rather good agreement with local realistic sources. They are
not necessarily representative of the entire Galaxy though. How potential differences
would impact the relative importance of $\phi_{\rm cat}$ with respect to the total flux
will be detailed in a forthcoming letter.

\subsection{Systematic errors from the catalog}

We have also plotted the uncertainty on $\phi_{\rm cat}$ that arises from the ages
and positions of the SNR of our catalog. These have been varied within the ranges
allowed by observations.
We did not consider the uncertainties on the ages and positions of pulsars.
The former are well determined. The rotation periods $P$ and drifts in time $\dot{P}$
of pulsars are actually measured with good accuracy.
The uncertainty on $\phi_\text{cat}$ corresponds to the grey bands of
Fig.~\ref{fig:confidence_local} and Fig.~\ref{fig:varianceSyst}.
The shape of these curves is explained by the fact that when $L$ is
lower (``min'' model), the effective range of diffusion in the plane
of the disk is reduced and the contribution of remote sources is
smaller. The relative contribution of the catalog is then larger for
smaller values of $L$.

\section{A mixed approach}

\begin{figure*}[h!]
\centerline{\includegraphics[width=8cm]{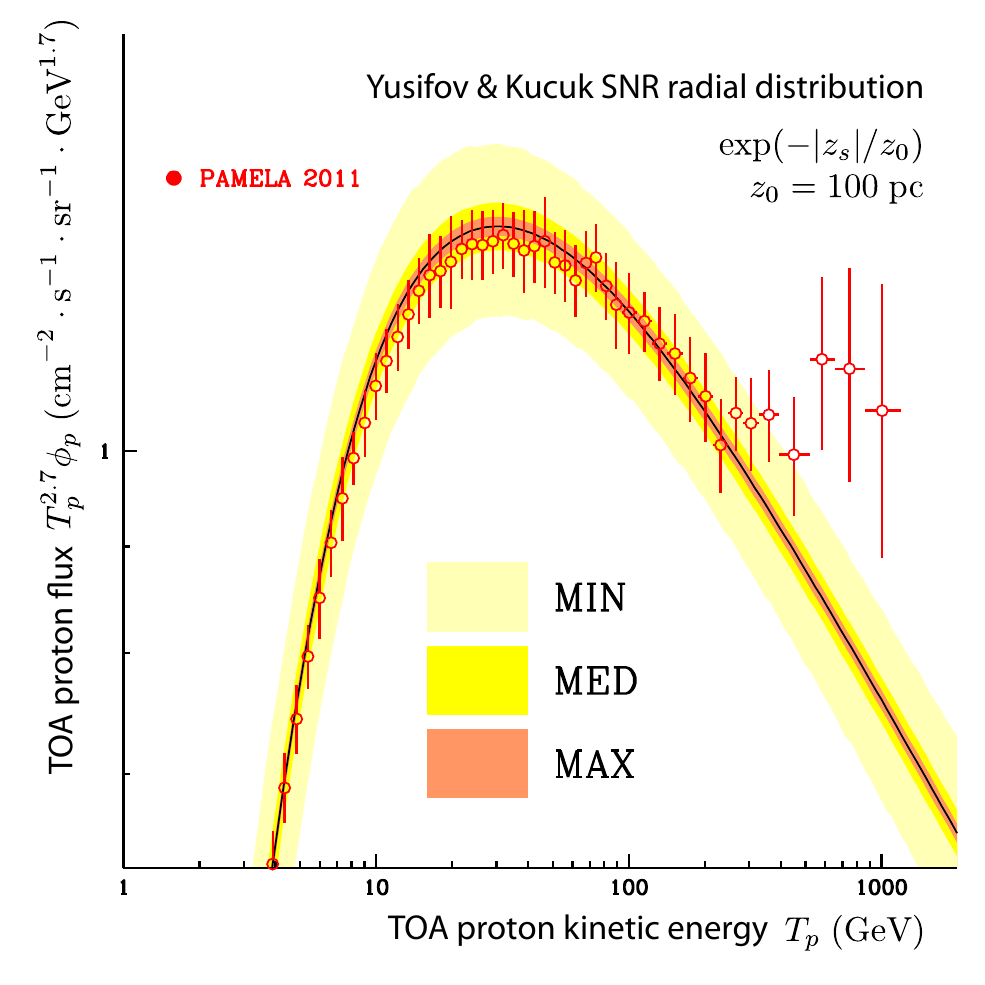}\includegraphics[width=8cm]{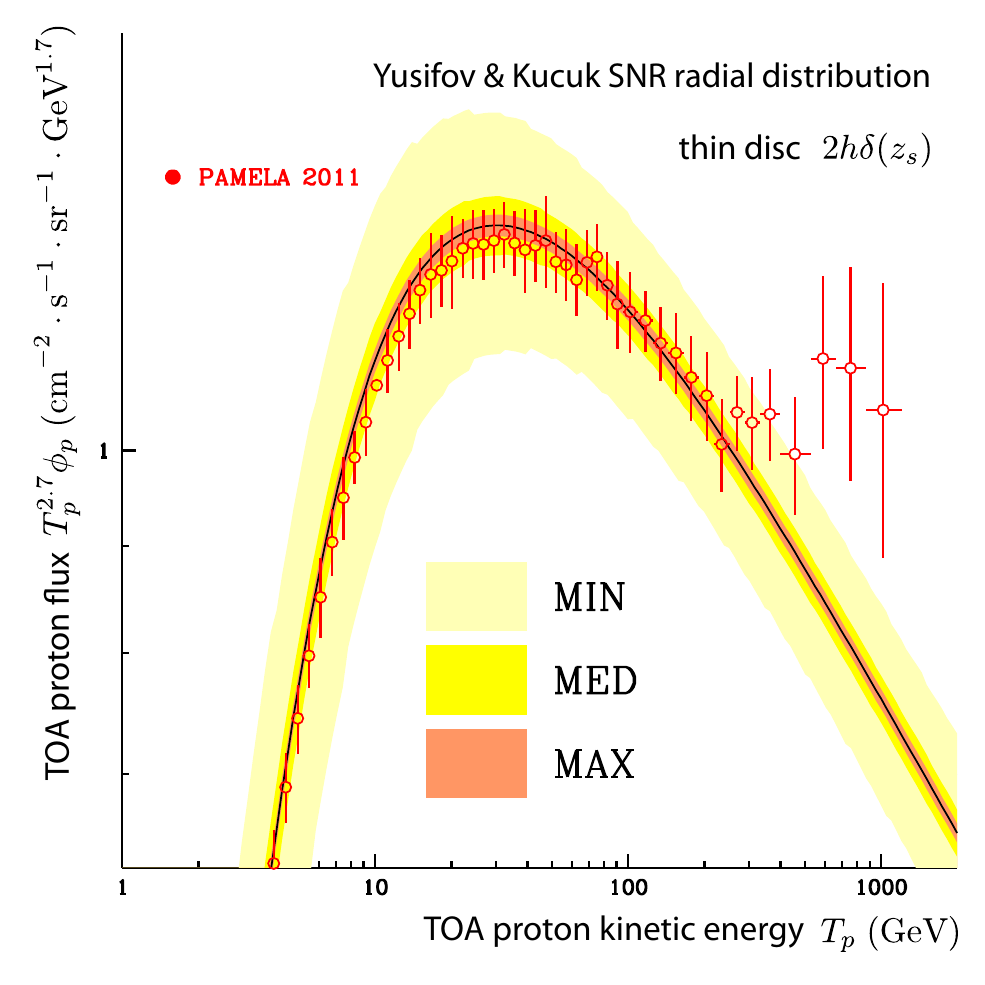}}
\caption{Mean flux (black curves) and
  envelopes representing the standard deviation of the flux, for the
  min, med and max propagation model and for
  an exponential distribution of sources along $z$ (left) and for a
  thin disk (right), for $\nu=3\;\text{century}^{-1}$.}
\label{fig:confidence_intervals}
\end{figure*}

The cosmic ray flux from the Galactic sources and the
associated uncertainties can be computed as 
$$\phi_\text{tot}(E) = \phi_\text{cat}(E) + \phi_\text{ext}(E)$$
where $\phi_\text{cat}$ is the flux coming from the sources belonging
to the catalog (closer than 2 kpc and younger than $3 \times 10^5$ yr) and
$\phi_\text{ext}(E)$ is the flux coming from sources which are
more distant than 2 kpc or older than $3 \times 10^5$ yr. 
The uncertainties on $\phi_\text{cat}(E)$ are obtained by
considering the observational errors on the source parameters (see
Fig.~\ref{fig:varianceSyst}). 
The uncertainty on $\phi_\text{ext}(E)$ can be evaluated by computing
the confidence intervals, as above. In some conditions, this is
equivalent to computing the usual the standard deviation of the flux,
as the variance of the flux located outside of the region covered by
the catalog is finite. This is shown in
Fig.~\ref{fig:confidence_intervals}. 

\section{Conclusions}

In the conventional model of Galactic CR propagation, the sources of primary
nuclei are described as a jelly which spans the disk of the Milky Way and
continuously injects particles inside the ISM. The actual distribution is
lacunary and consists of point-like objects which release cosmic rays in a
very short time. Because measurements have become very accurate, taking into
account the discreteness of the sources in our description of CR propagation
has become timely. In particular, the PAMELA~\citep{2011Sci...332...69A} and
CREAM~\citep{2011ApJ...728..122Y} observations point towards an excess in the
CR proton and helium fluxes with respect to the pure power-laws predicted
with the continuous model. Very few analyses have been devoted so far to the
myriad model. A challenging problem lies in the divergence of the variance
of the CR primary flux. As shown recently by~\citet{2011arXiv1105.4521B},
the second moment of the flux probability distribution function (PDF) is
infinite. Although the variance can always be regularized in some way or
another, this intriguing result threatens the conventional model of CR
propagation, so far extremely successful. That is why we have thoroughly
reinvestigated here how the discreteness of the sources may modify our
vision of Galactic CR propagation.

To commence, the CR flux of primary cosmic rays obtained in the myriad model
has a well-defined mean, identical to what is derived in the conventional
approach by averaging the source distribution over space and time. This good
point gives credit to the continuous model.
We have then concentrated on the issue of the flux variance which diverges
in the myriad model. Although the central limit theorem cannot be used in
that case, at least in its ordinary form, the PDF of the flux is well defined
everywhere inside the Galactic magnetic halo. We have run Monte Carlo
simulations with several thousands of different populations of point-like
sources. For the first time, we have derived the quantiles of the CR proton
flux as a function of proton energy. The quantiles turn out to be well-defined
and can be used to compute 68\% confidence intervals. In spite of an infinite
variance, meaningful error bars for the fluxes of primary CR nuclei can
be defined. These depend on the propagation parameters and tend to decrease
with the number of sources implied in the signal. That is why the uncertainty
bands of Fig.~\ref{fig:confidence_total} widen with the vertical extension
$L$ of the CR diffusive halo.

We have so far focused on the flux at the Earth, but the same procedure can
be applied throughout the Galaxy. It is however extremely time consuming and
we defer to a subsequent publication the analytical derivation of the flux
PDF. Such an investigation is crucial insofar as secondary species are produced
by the interactions of primary CR nuclei on the ISM and the actual distribution
of the latter matters. It would be interesting to examine the effect of the
myriad hypothesis on the fluxes of antiprotons or positrons at the Earth as
well as on the Galactic gamma ray diffuse emission. This emission has been
so far calculated in the framework of the continuous model. An essential
prediction is a gamma ray power law spectrum which traces the CR proton
and helium energy distributions. According to the continuous hypothesis, we
expect a gamma ray spectral index of $\sim - 2.8 \pm 0.1$ everywhere.
Measurements by the H.E.S.S. collaboration~\citep{Aharonian:2006au}
of the diffuse emission from the Galactic centre indicate a photon index of
$2.29 \pm 0.07 \pm 0.02$, significantly below the prediction of the continuous
model. Fluctuations of the CR proton and helium spectra around their mean values
have thus already been observed.

\begin{figure*}
\centering
\includegraphics[width=0.5\textwidth]{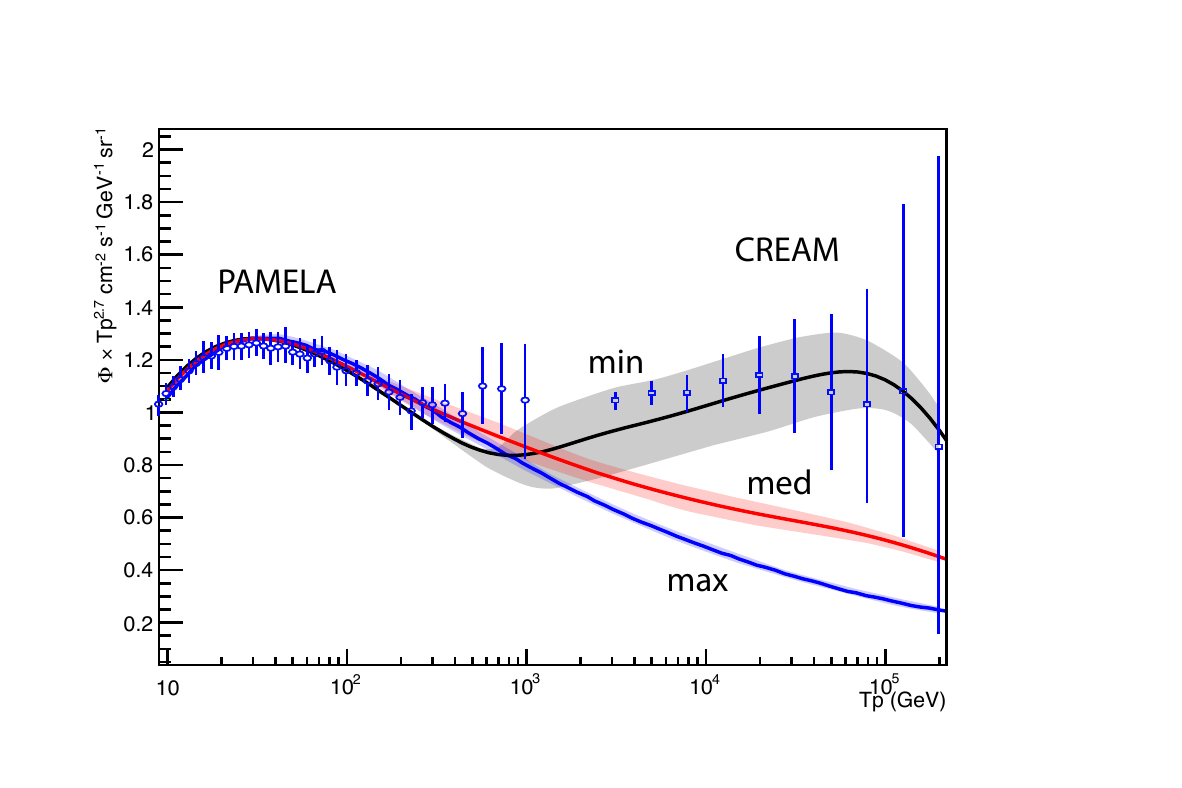}
\caption{
The blue, red and black curves feature the total flux $\phi$ computed as the
sum of the mean external flux $\langle \phi_{\rm ext} \rangle$ and the contribution
$\phi_{\rm cat}$ from the catalog. They respectively correspond to the max, med and
min CR propagation benchmark models. The bands which extend around the curves have
the same meaning as in Fig.~\ref{fig:confidence_local}. They indicate the standard
deviation of the flux associated to the observational errors on the ages and
distances of the SNR of the catalog.
}
\label{fig:varianceSyst}
\end{figure*}

As regards the primary CR fluxes at the Earth, another important result is the use
of a catalog for the nearby and young sources which span the so-called local region.
These objects have been extracted from the Green and ATNF surveys and are based
on the latest astronomical observations. The catalog yields a contribution
$\phi_{\rm cat}$ to the total flux $\phi$, whereas the sources extending beyond the
local region contribute the complement $\phi_{\rm ext}$. This procedure provides a
natural regularization of the flux variance since it is based on observations.
We have found that below a few hundreds of GeV, $\phi_{\rm cat}$ is always
negligible with respect to $\phi_{\rm ext}$, whatever the CR propagation
parameters. Since the flux is dominated by distant or old sources, the continuous
hypothesis applies and the use of the conventional CR propagation model is fully
justified. We have succeeded in reconciling the myriad model with the continuous
approach. Both should give identical results fo the low energy CR fluxes at the
Earth.
At the TeV scale and beyond, the situation becomes more complicated. If the total
number of objects that source the flux is large with respect to the catalog, the
contribution $\phi_{\rm ext}$ dominates and the continuous hypothesis is still
valid at high energies. On the contrary, if the supernova explosion rate $\nu$
becomes smaller than the canonical value of 3 events per century, or if the
half-thickness $L$ of the DH decreases, the catalog becomes relatively important
and may even dominate the flux above a few TeV. This is actually the case for the
black curve of Fig.~\ref{fig:varianceSyst} which corresponds to the min model and
$\nu = 1 \; \text{century}^{-1}$. The CREAM data points lie inside the catalog 68\%
error band. Notice how well the CR proton excess is naturally explained by local
and young sources which have actually been detected. There is no need for a break
in the injection spectrum, nor for a peculiar behaviour of the diffusion coefficient
$D$ with energy.
It should even be possible to find a particuliar set of CR propagation parameters
that would make the catalog a natural explanation of both PAMELA and CREAM data.
Since the local region alone is implied, these parameters ought not to be the same
as for the bulk of the Galactic magnetic halo.

\appendix

\section{Computation of the Green function}
\label{appendix:green}
\begin{figure*}[h!]
\centering
\includegraphics[width=0.5\textwidth]{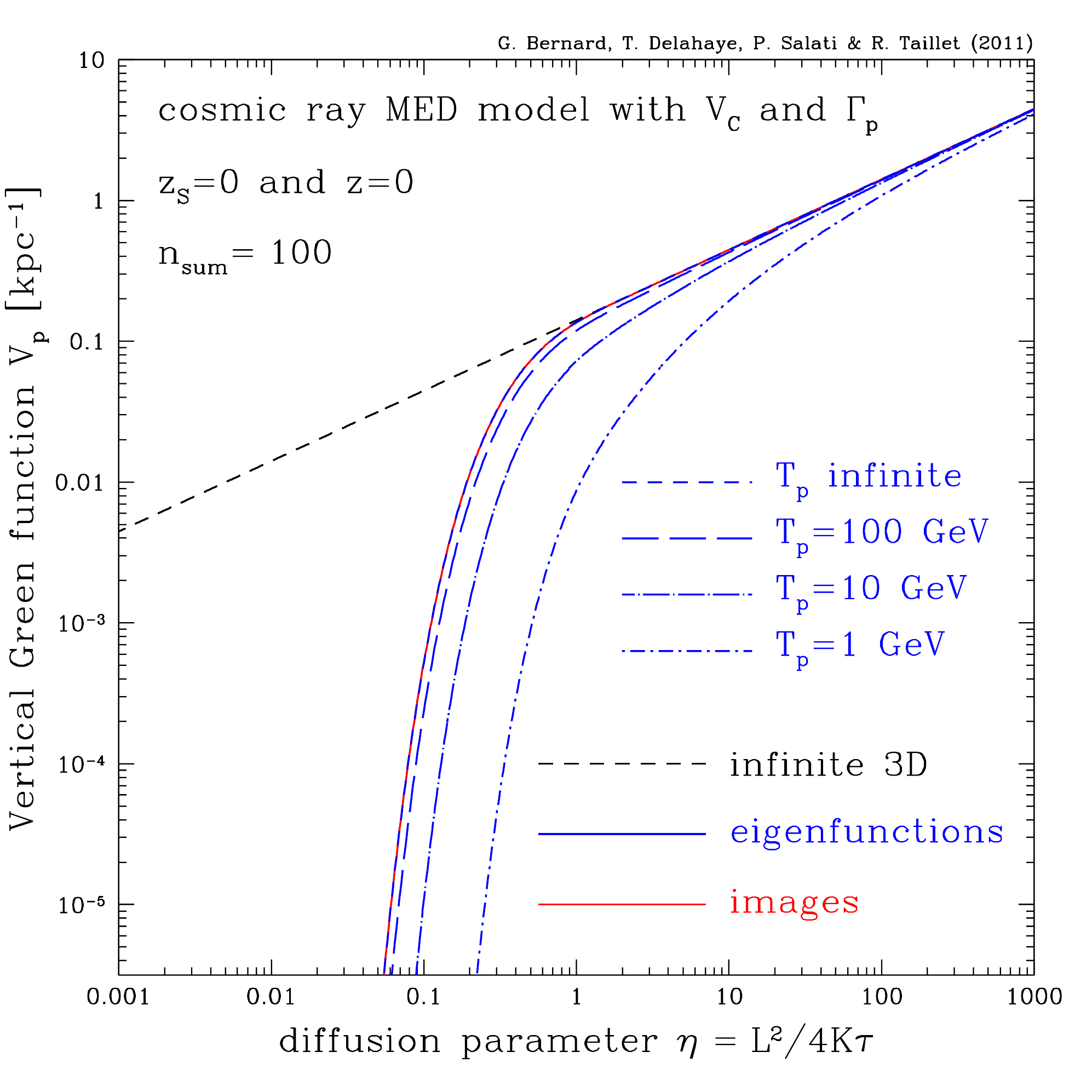}
\caption{
The effects of Galactic wind and spallations on the vertical
propagator $\mathscr{V}_{\! p}$ are presented in this figure where the MED
model of CR propagation, which best fits the B/C ratio, has been
selected. At very high proton energy, diffusion dominates over the other
processes and we recover the results obtained by setting $k_{w}$ and
$k_{s}$ equal to 0. Because $n_{\rm sum} = 100$ terms have been taken
into account in the expansions~(\ref{def:Vp}) and (\ref{V_image}), the
blue short dashed curve matches exactly the red solid line.
As the proton kinetic energy $T_{p}$ is decreased down to 1~GeV, Galactic
convection and proton interactions on the ISM become more and more
important and the blue curves departe from their infinite energy
limit.
}
\label{fig:V_p}
\end{figure*}

%
%
%
In order to solve analytically Eq.~(\ref{master_equation_G_p}), we
need to simplify our description of the Milky Way DH and replace it
by an infinite slab of half-thickness $L$ with a gaseous disk in the
middle at $z = 0$. Radial boundary conditions at $r = R_{\rm gal}$
are no longer implemented in the propagator. This simplification of
the setup for CR propagation could be a problem if we were interested
in the CR densities $\psi$ close to the radial boundaries, at a distance
of 20~kpc from the Galactic center. But our aim is to calculate these
densities at the Earth, at a galactocentric distance of
$r_{\odot} =$ 8.5~kpc, \textit{i.e.} far from the radial boundaries.
Furthermore, even though the propagator $\mathscr{G}_{\! p}$ is derived
within the framework of an infinite diffusive slab, integrals on
the sources of cosmic rays, such as relation~(\ref{psi_G_p}), are
still performed up to the radial boundaries at $r = R_{\rm gal}$.
We have checked that such a procedure does not introduce any
significant error on the CR fluxes at the Earth. In the case
of a source term $q_{\rm acc}$ that is continuous in space and time,
this approach yields results very close to those obtained with a method
based on radial Bessel functions.
With this simplified setup, the propagation of CR species becomes
invariant under a translation along the horizontal axes $x$ and $y$.
The master equation~(\ref{master_equation_G_p}) still needs to be
solved along the vertical direction $z$, with the condition that
$\mathscr{G}_{\! p}$ vanishes at the boundaries $z = \pm L$. The
construction of the Green function for CR nuclei is inspired from
the solution to the heat diffusion problem and has been given by many
authors.
The horizontal and vertical dependencies in $\mathscr{G}_{\! p}$ can be
factored out by setting
\begin{equation}
\mathscr{G}_{\! p} \left(
\mathbf{x} , t \, \leftarrow \, \mathbf{x}_{S} , t_{S}
\right) =
{\displaystyle \frac{1}{4 \pi D \tau}} \;
\exp \left( - \,
{\displaystyle \frac{\rho^{2}}{4 D \tau}} \right) \;
\mathscr{V}_{\! p} \left(
z , t \, \leftarrow \, z_{S} , t_{S} \right) \;\; ,
\label{def:G_p}
\end{equation}
where $\tau = t - t_{S}$ and
$\rho^{2} = (x - x_{S})^{2} \, + \, (y - y_{S})^{2}$.
The vertical function $\mathscr{V}_{\! p}$ is given by the expansion
\begin{equation}
\begin{split}
\mathscr{V}_p &(z , t \, \leftarrow  z_{S} , t_{S} )  =  \exp
\left\{ \frac{V_{C} \left( |z| - |z_{S}| \right)}{2 D} \right\}
\times \theta(\tau) \\
&\times 
\sum_{n=1}^{\infty}  \left\{
\frac{1}{C_n}  \, e^{ - \alpha_n  \tau}  \,
\mathscr{E}_n(z)  \, \mathscr{E}_n(z_{S})  
+  \frac{1}{C'_n} \, e^{ - \alpha'_n  \tau} \,
\mathscr{E}'_n(z) \, \mathscr{E}'_{n}(z_{S}) \right\} , \label{def:Vp} 
\end{split}
\end{equation}
where $\theta(\tau)$ is the Heaviside function. The sum runs over
even and odd eigenfunctions.
The former can be expressed as
\begin{equation}
\mathscr{E}_{n}(z) = \sin \left\{ k_{n} (L - |z|) \right\} \;\; .
\end{equation}
The corresponding even eigenwavevectors $k_{n}$ are solutions
of the equation
\begin{equation}
- \, \tan (k_{n} L) =
{\displaystyle \frac{k_{n}}{k_{w} + k_{s}}} \;\; ,
\end{equation}
where
$k_{w} \equiv {V_{C}}/{2 D}$ and
$k_{s} \equiv {h \Gamma_{p}}/{D}$ are typical wavevectors which account
for the effects of Galactic convection and proton collisions on the ISM.
Notice that at high energy, the dimensionless parameters $k_{w} L$ and
$k_{s} L$ become very small as $D$ increases. We then expect convection
and spallations to have little effect with respect to space diffusion
at high energy, as illustrated in Fig.~\ref{fig:V_p}.
The odd eigenfunctions are given by
\begin{equation}
\mathscr{E}'_{n}(z) = \sin \left\{ k'_{n} (L - z) \right\} \;\; ,
\end{equation}
where the odd eigenwavevectors $k'_{n}$ are such that
$k'_{n} L = n \pi$. This condition ensures that $\mathscr{E}'_{n}(z)$
vanishes at $z = 0$ and is an odd function of the height $z$.
The decay rates $\alpha_{n}$ and $\alpha'_{n}$ are respectively
related to the eigenwavevectors $k_{n}$ and $k'_{n}$ through
\begin{equation}
\alpha_{n} \, = \, D k_{n}^{2} \, + \,
{\displaystyle \frac{V_{C}^{2}}{4 D}} \;\; .
\end{equation}
\begin{equation}
\alpha_{n'} \, = \, D k_{n'}^{2} \, + \,
{\displaystyle \frac{V_{C}^{2}}{4 D}} \;\; . \nonumber
\end{equation}
The eigenfunctions $\mathscr{E}_{n}(z)$ and $\mathscr{E}'_{n}(z)$ may be
understood as orthogonal vectors of a basis over which any vertical
function that vanishes at $z = \pm L$ can be expanded. They are
normalized in such a way that
\begin{equation}
\langle \mathscr{E}_{p} \, | \, \mathscr{E}_{n} \rangle \equiv
{\displaystyle \int_{-L}^{+L}} \, dz \; \mathscr{E}^{\star}_{p}(z) \;
\mathscr{E}_{n}(z) = \delta_{pn} \, C_{n}
\end{equation}
and
\begin{equation}
\langle \mathscr{E}'_{p} \, | \, \mathscr{E}'_{n} \rangle =
\delta_{pn} \, C'_{n} \equiv \delta_{pn} \, L \;\; ,
\label{eq:ortho_even_odd}
\end{equation}
where the normalization factors $C_{n}$ and $C'_{n}$ are
\begin{equation}
C_{n} = L \, \left\{ 1 \, + \,
{\displaystyle \frac{{\rm sinc}^{2}(k_{n}L)}{p}} \right\}
\;\;\;{\rm and}\;\;\;
C'_{n} = L \;\; .
\end{equation}
The dimensionless parameter $p$ is defined by
${1}/{p} = (k_{w} + k_{s}) L$.
It becomes infinite in the absence of Galactic convection and collisions
on the ISM. In that limit, the eigenwavevectors $k_{n}$ are equal to
${(n - 1/2) \, \pi}/{L}$. 
%
%

%
%
%
We still need to address a technical problem related to the behaviour
of the vertical propagator~(\ref{def:Vp}) when $t$ is close to $t_{S}$.
We shall find that the variance of the CR proton flux is dominated by
local and recent supernova explosions. In that case, the period of time
$\tau$ that separates the injection of protons from their detection at
the Earth becomes small. The expansion~(\ref{def:Vp}) needs to be pushed
a long way up in terms of the number of eigenfunctions involved. When
$\tau$ is small, the exponentially decreasing functions
$\exp(- \alpha_{n} \tau)$ and
$\exp(- \alpha'_{n} \tau)$ are still close to unity unless $n$ becomes
very large. The numerical convergence of $\mathscr{V}_{\! p}$ requires then
to sum expression~(\ref{def:Vp}) over an exceedingly large number
of terms, hence a potential problem of CPU time.
Because in that regime, most of the expansion~(\ref{def:Vp}) is provided
by high order terms, Galactic convection and CR spallations become
negligible with respect to diffusion. For these terms, the even
eigenwavevectors $k_{n}$ are actually very close to their pure diffusion
values of ${(n - 1/2) \, \pi}/{L}$, even in the case where $p$ is not
very large. In this regime of basically pure diffusion, another solution
to CR propagation is provided by the method of electrical images.
To commence, if the half-thickness $L$ of the slab is made infinite, we
recover pure diffusion in infinite 3D space. In that case, the propagator
is well-known, with a vertical contribution expressed as
\begin{equation}
\begin{split}
\mathscr{V}_{\! p}
\left( z , t \, \leftarrow \, z_{S} , t_{S} \right)
&\equiv
\mathscr{V}_{\rm 1D}
\left( z , t \, \leftarrow \, z_{S} , t_{S} \right) \\
& = \frac {\theta(\tau)}{\sqrt{4 \pi D \tau}} \,
\exp \left\{ - 
\frac{\left( z - z_{S} \right)^{2}}{4 D \tau} \right\} \;\; .
\label{propagator_reduced_1D}
\end{split}
\end{equation}
As already discussed by~\citet{1999PhRvD..59b3511B}, the vertical boundaries
of the DH can now be implemented by associating to each point-like source
lying at height $z_{S}$ the infinite series of its multiple images through
the planes $z = \pm L$. The boundaries act as mirrors and give from the source
an infinite series of images. The n-th image is located at
\begin{equation}
z_{n} \, = \,
2 L \, n \, + \, \left( -1 \right)^{n} z_{S} \;\; ,
\end{equation}
and has a positive or negative contribution depending on whether $n$
is an even or odd number. When the diffusion time $\tau = t - t_{S}$
is small, the 1D solution~(\ref{propagator_reduced_1D}) is a quite good
approximation. The relevant parameter is actually
\begin{equation}
\eta \, = \,
{\displaystyle \frac{L^{2}}{4 D \tau}} \;\; ,
\label{definition_eta}
\end{equation}
and, in the regime where it is much larger than 1, \textit{i.e.}, for small
values of $\tau$, the propagation is insensitive to the vertical
boundaries. When the diffusion parameter $\eta$ decreases and the
diffusion length $\lambda_{\rm D} \equiv \sqrt{4 D \tau}$ becomes
comparable to the half-thickness $L$ of the DH, images of the point-like
source need to be taken into account in the sum
\begin{equation}
\mathscr{V}_{\! p}
\left( z , t \, \leftarrow \, z_{S} , t_{S} \right) \, = \,
{\displaystyle \sum_{n \, = \, - \infty}^{+ \infty}} \,
\left( -1 \right)^{n} \;
\mathscr{V}_{\rm 1D}
\left( z , t \, \leftarrow \, z_{n} , t_{S} \right) \;\; .
\label{V_image}
\end{equation}
When $\eta$ is much smaller than 1, many terms must be taken into account
in the previous sum. However, this regime corresponds to large values of
the diffusion time $\tau$ for which expansion~(\ref{def:Vp}) converges
very rapidly.

%
%
%
We have devised two complementary methods to calculate the propagator
$\mathscr{G}_{\! p}$. Depending on the value of the diffusion parameter
$\eta$, we can use the relation~(\ref{V_image}) of electrical images
($\eta \geq 1$) or the expansion~(\ref{def:Vp}) ($\eta \leq 1$).
Since the CR sources and the Earth are located inside the Galactic
disk, we have set $z_{S}$ and $z$ equal to 0 in Fig.~\ref{fig:V_p}.
The vertical part of the Green function depends only on the time
delay $\tau = t - t_{S}$ with
$\mathscr{V}_{\! p}(\tau) \equiv
\mathscr{V}_{\! p} (0 , t    \, \leftarrow \, 0 , t_{S}) =
\mathscr{V}_{\! p} (0 , \tau \, \leftarrow \, 0 , 0    )$.
At low energy, Galactic convection and CR collisions on the ISM
become important with respect to space diffusion. The method of
electrical images is reliable only for small values of the diffusion
time $\tau$. Expansion~(\ref{V_image}) can still be used to calculate
$\mathscr{V}_{\! p}$ in the regime where $\eta$ becomes large. How large depends
on the relative strength of the various CR propagation mechanisms. To get
a flavor of the range of validity over which electrical images can be used
even in the case where $k_{s}$ and $k_{w}$ are larger than ${1}/{L}$, we have
borrowed the MED set of CR propagation parameters from~\citet{2004PhRvD..69f3501D}.
This benchmark configuration provides the best fit to the B/C
measurements (see Table~1).
The black short dashed line of Fig.~\ref{fig:V_p} corresponds to pure
diffusion in infinite 3D space.
The method of electrical images can only be used for pure CR diffusion
and yields the red solid line. Expansion~(\ref{V_image}) has been
calculated with $n_{\rm sum} = 100$ images and is valid down to the small
value of $\eta \sim 0.016$.
Expansion~(\ref{def:Vp}) has also been pushed up to the 100-th term and
converges even for $\eta$ as large as $10^{3}$. It leads to the blue
curves which correspond each to a different proton kinetic energy $T_{p}$.
Because diffusion takes over the other processes at very high energy,
the blue short dashed curve is completely superimposed on the red solid
line. In that regime, expansions~(\ref{def:Vp}) and (\ref{V_image}) yield
the same result.
As $T_{p}$ decreases, convection and spallations come into play and the
blue curves depart from their high energy limit. Notice that the
100~GeV configuration is still fairly close to the high energy case.
At 10~GeV, the blue dotted-long dashed curve differs noticeably from
the red solid line. The pure diffusive regime is nevertheless obtained for
$\eta \geq 100$. Finally, the 1~GeV blue dotted-short dashed curve is
significantly shifted towards higher values of $\eta$ with respect to the
pure diffusive case. In order to reliably calculate the proton propagator
below $\sim$ 1~GeV, many terms need to be taken into account in
expansion~(\ref{def:Vp}). This sum should be used up to large values of
$\eta$ before the electrical images provide an accurate result.

\section{Probability distribution for the flux}
\label{appendix:proba}

In this section, we compute the high-$\phi$ behaviour of the
probability density $P(\phi)$ for the flux due to a point source drawn
from a uniform spatial and temporal distribution. This high-$\phi$
tail of the distribution is the part leading to the divergence of the
variance. It is due to the sources which are very close and very
young, for which the effects of Galactic wind, escape and
reacceleration are very small. We can neglect these effects, as long
as we are only interested in the asymptotic behaviour of $P(\phi)$ at
high $\phi$.

The flux at distance $r$ and at time $t$ from a point source emitting
instantly all its particles at $r=0$ and $t=0$ is given by
$$\phi = \frac{a}{t^{3/2}} e^{-r^2/4Kt} \quad
\text{where} \quad a = \frac{q}{(4\pi K)^{3/2}}$$

\subsection*{Case of a 2D distribution of sources (thin disk)}

Let us first consider sources having a given age $t$. The probability density
that a source lies at a distance $r$ is
given by $$p(r) \equiv \frac{dP(r)}{dr} = \frac{2r}{R^2}$$
where $R$ stands for the radius of the region containing the sources.
We have
$$d\phi = -\frac{2r\, dr}{4Kt} \, \phi$$
so that
$$p(\phi | t) = \frac{dP(\phi,t)}{d\phi} = \frac{4Kt}{R^2} \, \frac{1}{\phi}$$
For an age $t$, fluxes are in the interval 
$$\frac{a}{t^{3/2}} e^{-R^2/4Kt} \leq \phi \leq \frac{a}{t^{3/2}}$$
which can be written as
$$p(\phi | t) = \frac{4Kt}{R^2} \, \frac{1}{\phi} 
W\left(\frac{a}{t^{3/2}} e^{-R^2/4Kt} ,
\frac{a}{t^{3/2}} \right)$$
Where $W$ stands for the window function, being equal to 1 in the
interval and 0 outside.
It is easily checked that
$$\int p(\phi | t) d\phi = \frac{4Kt}{R^2} \ln
\frac{\phi_\text{max}}{\phi_\text{min}} = 1$$

The probability distribution for $\phi$ is obtained by
$$p(\phi) = \int p(\phi | t) \, p(t) \, dt$$
where the upper bound is
$$t_\text{max} = a^{2/3} \phi^{-2/3}$$
the lower bound $t_\text{min}(\phi)$ is a solution of
$$\phi = \frac{a}{t_\text{min}^{3/2}} e^{-R^2/4Kt_\text{min}}$$ 
If the age distribution $p(t)$ is uniform between $0$ and $T$, we have
$p(t)=1/T$ so that
$$p(\phi) = \frac{4K}{R^2T \phi}\int_{t_\text{min}}^{a^{2/3}  \phi^{-2/3}}  \, t\, dt$$
$$p(\phi) = \frac{2K}{R^2T \phi} \left[ a^{4/3} \phi^{-4/3} - t_\text{min}^2(\phi)\right]$$
$$p(\phi) = \frac{2Ka^{4/3}}{R^2T} \left[  \phi^{-7/3} -
  \frac{t_\text{min}^2(\phi)}{\phi a^{4/3}}\right]$$
This can be written as
$$p(\phi) = \frac{2Ka^{4/3}}{R^2T} \phi^{-7/3} \left[ 1 - 
e^{-R^2/4Kt_\text{min}(\phi)} \right] $$
At high $\phi$, we have
$$p(\phi) \propto \phi^{-7/3}$$

\subsection*{Case of a 3D distribution of sources (thick disk)}

If the sources are distributed in a volume instead of a surface, we
still have
$$d\phi = -\frac{r\, dr}{2Kt} \, \phi$$
but
$$p(r) \equiv \frac{dP(r)}{dr} = \frac{3r^2}{R^3}$$
$$dP = \frac{3r^2 dr}{R^3} = \frac{6 r Kt}{R^3} \frac{d\phi}{\phi}$$
As
$$r= \sqrt{-4Kt \ln (\phi t^{3/2}/a)}$$
we have
$$p(\phi | t) = \frac{12(Kt)^{3/2}}{R^3}\frac{\sqrt{-\ln(\phi
    t^{3/2}/a)}}{\phi}$$
As before, we can check that the total probability is 1. Let us compute
$$\int p(\phi | t) \, d\phi = \frac{12(Kt)^{3/2}}{R^3} \int \frac{\sqrt{-\ln(\phi
    t^{3/2}/a)}}{\phi} \, d\phi$$
We set $x=\phi t^{3/2}/a$,
$$\int p(\phi | t) \, d\phi = \frac{12(Kt)^{3/2}}{R^3} \int_{\phi_\text{min} t^{3/2}/a}^{\phi_\text{max} t^{3/2}/a}
\frac{\sqrt{- \ln x}}{x} \, dx$$
The integral is given by
$$\int \frac{\sqrt{-\ln x}}{x} \, dx = \frac{2}{3} \left( -\ln x \right)^{3/2} + \text{cte}$$
so that
$$\int p(\phi | t) \, d\phi = \frac{8(Kt)^{3/2}}{R^3} \left[ \left(
  -\ln x \right)^{3/2}
  \right]_{\phi_\text{min} t^{3/2}/a}^{\phi_\text{max} t^{3/2}/a}$$
Finally, as $\phi_\text{max} t^{3/2} = a$ and $\phi_\text{min} t^{3/2}= a e^{-R^2/4Kt}$, 
$$\int p(\phi | t) \, d\phi = \frac{8(Kt)^{3/2}}{R^3} \left(
  \frac{R^2}{4Kt} \right)^{3/2}=1$$

The probability distribution is obtained by
$$p(\phi) = \int p(\phi | t) \, p(t) \, dt$$
As before, $p(t) = 1/T$,
$$p(\phi) = \int \frac{12(Kt)^{3/2}}{TR^3} \frac{\sqrt{-\ln(\phi
    t^{3/2}/a)}}{\phi} dt$$
where the bounds $t_\text{min}$ and  $t_\text{max}$ are the same as
before. We define $y=t\phi^{2/3}/a^{2/3}$,
$$p(\phi) = \int \frac{12a^{5/3}(Ky)^{3/2}}{TR^3\phi} \frac{\sqrt{-\ln y^{3/2}}}{\phi} \frac{dy}{\phi^{2/3}}$$
$$p(\phi) = \frac{12a^{5/3}K^{3/2}}{TR^3\phi^{8/3}}  \sqrt{\frac{3}{2}}\int y^{3/2}\sqrt{-\ln y} \, dy$$
with $y_\text{max} = 1$ and $y_\text{min}(\phi)$ solution of
$y= \exp(-R^2a\phi^{2/3}/6Ky)$. For high values of $\phi$, the lower
bound vanishes and the integral does not depend on $\phi$, which
yields the final result
$$p(\phi) \propto \phi^{-8/3}$$

\begin{acknowledgements}
We thank Prof. Blasi for very useful discussions. This work was supported by the Spanish MICINN's  Consolider-Ingenio 2010 Programme under grant CPAN CSD2007-00042. We also thank the support of the MICINN under grant FPA2009-08958, the Community of Madrid under grant HEPHACOS S2009/ESP-1473,and the European Union under the Marie Curie-ITN program PITN-GA-2009-237920
\end{acknowledgements}

\bibliographystyle{aa} 
\bibliography{papier_variance}

\begin{thebibliography}{18}
\expandafter\ifx\csname natexlab\endcsname\relax\def\natexlab#1{#1}\fi

\bibitem[{{Adriani} {et~al.}(2011){Adriani}, {Barbarino}, {Bazilevskaya},
  {Bellotti}, {Boezio}, {Bogomolov}, {Bonechi}, {Bongi}, {Bonvicini},
  {Borisov}, {Bottai}, {Bruno}, {Cafagna}, {Campana}, {Carbone}, {Carlson},
  {Casolino}, {Castellini}, {Consiglio}, {De Pascale}, {De Santis}, {De
  Simone}, {Di Felice}, {Galper}, {Gillard}, {Grishantseva}, {Jerse},
  {Karelin}, {Koldashov}, {Krutkov}, {Kvashnin}, {Leonov}, {Malakhov},
  {Malvezzi}, {Marcelli}, {Mayorov}, {Menn}, {Mikhailov}, {Mocchiutti},
  {Monaco}, {Mori}, {Nikonov}, {Osteria}, {Palma}, {Papini}, {Pearce},
  {Picozza}, {Pizzolotto}, {Ricci}, {Ricciarini}, {Rossetto}, {Sarkar},
  {Simon}, {Sparvoli}, {Spillantini}, {Stozhkov}, {Vacchi}, {Vannuccini},
  {Vasilyev}, {Voronov}, {Yurkin}, {Wu}, {Zampa}, {Zampa}, \&
  {Zverev}}]{2011Sci...332...69A}
{Adriani}, O., {Barbarino}, G.~C., {Bazilevskaya}, G.~A., {et~al.} 2011,
  Science, 332, 69

\bibitem[{Aharonian {et~al.}(2006)}]{Aharonian:2006au}
Aharonian, F. {et~al.} 2006, Nature, 439, 695

\bibitem[{{Baltz} \& {Edsj{\"o}}(1999)}]{1999PhRvD..59b3511B}
{Baltz}, E.~A. \& {Edsj{\"o}}, J. 1999, \prd, 59, 023511

\bibitem[{{Blasi} \& {Amato}(2011)}]{2011arXiv1105.4521B}
{Blasi}, P. \& {Amato}, E. 2011, ArXiv e-prints

\bibitem[{{Delahaye} {et~al.}(2010){Delahaye}, {Lavalle}, {Lineros}, {Donato},
  \& {Fornengo}}]{2010A&A...524A..51D}
{Delahaye}, T., {Lavalle}, J., {Lineros}, R., {Donato}, F., \& {Fornengo}, N.
  2010, \aap, 524, A51

\bibitem[{{Donato} {et~al.}(2004){Donato}, {Fornengo}, {Maurin}, {Salati}, \&
  {Taillet}}]{2004PhRvD..69f3501D}
{Donato}, F., {Fornengo}, N., {Maurin}, D., {Salati}, P., \& {Taillet}, R.
  2004, \prd, 69, 063501

\bibitem[{{Donato} {et~al.}(2001){Donato}, {Maurin}, {Salati}, {Barrau},
  {Boudoul}, \& {Taillet}}]{2001ApJ...563..172D}
{Donato}, F., {Maurin}, D., {Salati}, P., {et~al.} 2001, \apj, 563, 172

\bibitem[{{Gradshteyn} {et~al.}(2007){Gradshteyn}, {Ryzhik}, {Jeffrey}, \&
  {Zwillinger}}]{2007tisp.book.....G}
{Gradshteyn}, I.~S., {Ryzhik}, I.~M., {Jeffrey}, A., \& {Zwillinger}, D. 2007,
  {Table of Integrals, Series, and Products}, ed. {Gradshteyn, I.~S., Ryzhik,
  I.~M., Jeffrey, A., \& Zwillinger, D. }

\bibitem[{Green(2009)}]{Green2009}
Green, D.~A. 2009, A Catalogue of Galactic Supernova Remnants (2009 March
  version) (Cambridge)

\bibitem[{{Higdon} {et~al.}(2003){Higdon}, {Higdon}, {van der Hulst}, \&
  {Stacey}}]{2003ApJ...592..161H}
{Higdon}, S.~J.~U., {Higdon}, J.~L., {van der Hulst}, J.~M., \& {Stacey}, G.~J.
  2003, \apj, 592, 161

\bibitem[{{Manchester} {et~al.}(2005){Manchester}, {Hobbs}, {Teoh}, \&
  {Hobbs}}]{2005AJ....129.1993M}
{Manchester}, R.~N., {Hobbs}, G.~B., {Teoh}, A., \& {Hobbs}, M. 2005, \aj, 129,
  1993

\bibitem[{{Maurin} {et~al.}(2001){Maurin}, {Donato}, {Taillet}, \&
  {Salati}}]{2001ApJ...555..585M}
{Maurin}, D., {Donato}, F., {Taillet}, R., \& {Salati}, P. 2001, \apj, 555, 585

\bibitem[{{Nakamura et al.}(2010)}]{Nakamura}
{Nakamura et al.}, K. 2010, J. Phys. G: Nucl. Part. Phys., 37, 075021

\bibitem[{{Norbury} \& {Townsend}(2007)}]{2007NIMPB.254..187N}
{Norbury}, J.~W. \& {Townsend}, L.~W. 2007, Nuclear Instruments and Methods in
  Physics Research B, 254, 187

\bibitem[{{Taillet} {et~al.}(2004){Taillet}, {Salati}, {Maurin},
  {Vangioni-Flam}, \& {Cass{\'e}}}]{2004ApJ...609..173T}
{Taillet}, R., {Salati}, P., {Maurin}, D., {Vangioni-Flam}, E., \& {Cass{\'e}},
  M. 2004, \apj, 609, 173

\bibitem[{{Yoon} {et~al.}(2011){Yoon}, {Ahn}, {Allison}, {Bagliesi}, {Beatty},
  {Bigongiari}, {Boyle}, {Childers}, {Conklin}, {Coutu}, {DuVernois}, {Ganel},
  {Han}, {Jeon}, {Kim}, {Lee}, {Lutz}, {Maestro}, {Malinine}, {Marrocchesi},
  {Minnick}, {Mognet}, {Nam}, {Nutter}, {Park}, {Park}, {Seo}, {Sina},
  {Swordy}, {Wakely}, {Wu}, {Yang}, {Zei}, \& {Zinn}}]{2011ApJ...728..122Y}
{Yoon}, Y.~S., {Ahn}, H.~S., {Allison}, P.~S., {et~al.} 2011, \apj, 728, 122

\bibitem[{{Yuan} {et~al.}(2011){Yuan}, {Zhang}, \& {Bi}}]{2011PhRvD..84d3002Y}
{Yuan}, Q., {Zhang}, B., \& {Bi}, X.-J. 2011, \prd, 84, 043002

\bibitem[{{Yusifov} \& {K{\"u}{\c c}{\"u}k}(2004)}]{2004A&A...422..545Y}
{Yusifov}, I. \& {K{\"u}{\c c}{\"u}k}, I. 2004, \aap, 422, 545

\end{thebibliography}

\end{document}